\documentclass[HFI]{baltzer}
\usepackage{epsfig}
\begin{document}
\begin{frontmatter}
%
\title{Nuclear Structure Studies at ISOLDE and their Impact on the 
Astrophysical r-Process}       
\author[A]{Karl-Ludwig Kratz}      
\author[A]{Bernd Pfeiffer}      
\author[B]{Friedrich-Karl Thielemann}
\author[C]{William B. Walters}
\address[A]{Institut f\"ur Kernchemie, Universit\"at Mainz, Germany 
\email{klkratz@vkcmzd.chemie.uni-mainz.de}}
\address[B]{Departement f\"ur Physik und Astronomie, Universit\"at Basel, 
Switzerland
\email{fkt@quasar.physik.unibas.ch}}
\address[C]{Department of Chemistry, University of Maryland, USA
\email{ww3@umail.umd.edu}}
\runningauthor{Kratz et al.}
\runningtitle{Astrophysics at ISOLDE}
%
\begin{abstract}  
The focus of the present review is the production of the heaviest elements
in nature via the r-process. A correct understanding and modeling requires 
the knowledge of nuclear properties far from stability and a detailed
prescription of the astrophysical environment. Experiments at CERN/ISOLDE have
played a pioneering role in exploring the characteristics of nuclear
structure in terms of masses and $\beta$-decay properties. Initial examinations 
paid attention to far unstable nuclei with magic neutron numbers related to 
r-process peaks, while present activities are centered on
the evolution of shell effects with the distance from the valley of stability.
We first show in site-independent applications the effect of both types of 
nuclear properties on r-process abundances. Then, we explore the results of
calculations related to two different `realistic' astrophysical sites,
(i) the supernova neutrino wind and (ii) neutron star mergers.
We close with a list of remaining theoretical and experimental challenges 
needed to overcome for a full understanding of the nature of the r-process, 
and the role CERN/ISOLDE can play in this process.
\end{abstract}
\end{frontmatter}
\section{Introduction}

The astrophysical rapid neutron-capture process (r-process) has been 
recognized for a long time as the scenario responsible for the synthesis
of approximately half of the nuclear species in nature,
which are more massive than Fe \cite{b2fh,cameron57}.
It requires environments
with a high neutron density, where neutron captures are faster 
than $\beta$-decays, even for neutron-rich unstable nuclei up to 15-30 units 
from stability.  Only under such conditions it is possible that 
highly unstable nuclei are produced near the neutron drip-line 
via neutron captures, $(\gamma,n)$-photodisintegrations, $\beta^-$-decays
and $\beta$-delayed processes, leading also to the formation of the heaviest
elements in nature like Th, U, and Pu. Far from stability,
magic neutron numbers are encountered for smaller mass
numbers A than in the valley of stability, which shifts the r-process
abundance peaks in comparison to the s-process peaks (which occur at neutron 
shell closures at the stability line due to small neutron
capture cross sections). 
Besides this basic understanding, the history of r-process research has
been quite diverse in suggested scenarios (for reviews see 
\cite{seeger65,cameron70,hillebrandt78,mathward85,cowtt91,thiel98,wallerstein97,kaeppeler98}).
Starting with a seed distribution somewhere around
A=50-80 before massive neutron-capture sets in, the operation of an r-process 
requires 10 to 150 neutrons per seed nucleus to form all heavier r-nuclei.
The question is which kind of environment can provide such a supply of 
neutrons to act before decaying with a 10 min half-life.
The logical conclusion is that only explosive environments, producing
or releasing these neutrons suddenly, can account for such conditions.
Without having discussed any stellar models yet, we want to mention some
recent literature on these issues.
The r-process site is not a settled one; however, two astrophysical
settings are suggested most frequently: (i) Type II supernovae (SNII) with 
postulated high-entropy ejecta 
\cite{woosley94b,takahashi94,freiburghaus99},
where the delayed emergence of r-process matter in galactic evolution
\cite{mcwilliam95,mcwilliam97} indicates that these can probably only
be SN with small progenitor mass; and (ii) neutron star mergers or other 
events with low-entropy ejecta 
\cite{lattimer77,meyer89,eichler89,rosswog99a}.

\subsection{Equilibria in Explosive Burning}

After this ad hoc introduction of sites, we want to return first
to the conditions in high-temperature explosive plasmas in order to
test what kind of nuclear-physics input is required, and also what type of
thermodynamic environment properties can lead to the neutron to seed
ratios mentioned above. The main aspect is that
a whole variety of burning processes responsible for the abundances
of intermediate and heavy nuclei, like e.g. explosive O and Si-burning, are 
leading to partial (quasi) or full equilibria of reactions 
(QSE or NSE). These are described by the chemical potentials of nuclei
which form a Boltzmann gas. The abundance ratios are (besides thermodynamic
environment properties) determined by mass differences. Therefore, mass
uncertainties matter, while uncertainties in cross sections
do not enter abundance determinations.
Non-equilbrium regions are identified by small reaction cross
sections, either due to small Q-values for reactions
out of the magic numbers or due to small level or resonance
densities for light nuclei. For sufficiently high temperatures all QSE groups
merge to a full NSE.

The chemical equilibrium for neutron or proton captures leads to abundance
maxima at specific neutron or proton separation energies

\begin{eqnarray}
{{Y(A_c)}\over {Y(A_t)}}&=&n_{p} {{G(A_c)}\over{g_pG(A_t)}}
\Bigl[{{A_c}\over {A_pA_t}}\Bigr]^{3/2} \Bigl[{{2\pi
\hbar^2}\over {m_ukT}}\Bigr]^{3/2} {\rm exp}(S_{p}(A_c)/kT) \\
{S_{p}(A_c) \over kT}&=& 24 \ln\Bigl[
({A_tA_p \over {A_c}})^{3/2}
({G(A_t)g_p \over 2G(A_c)})^{3/2}
({T \over {10^9{\rm K}}})^{3/2}
{N_A \over {n_{p}/{\rm cm}^3}}\Bigr].
\end{eqnarray}

These equations are valid for neutron, proton and/or $\alpha$ captures, as 
indicated by the subscript $p$ (projectile); subscript $t$ stands for target 
and $c$ for the compound
nucleus. The $Y$'s are abundances related to number densities
$n$ via $n=\rho N_A Y$, where $\rho$ denotes the mass density and $N_A$
the Avogadro number. At the maximum in an isotopic or isotonic
line we have $S_{p=n {\rm\ or\ } p}\approx 24kT$, if
the partition functions G are neglected. This is slightly 
modified by logarithmic dependences on projectile (neutron or proton)
density $n_p$ and temperature $T$. In case
an equilibrium with neutrons and protons exists, the abundance
maximum is found in the nuclear chart at the intersection of the relevant
neutron and proton separation energies. The free neutron and proton densities
reflect the total neutron/proton ratio in matter, which is determined by slow
(not in equilibrium), weak interactions via changing the total proton/nucleon 
ratio $Y_e=$$<$$Z/A$$>$.

The understanding that explosive burning stages are governed by QSE
or even NSE is growing \cite{woosley73,hixthi96,hixthi99,wallerstein97}.
This has been shown recently in calculations of
SNII nucleosynthesis \cite{hoffman99,thielemann96a}
with two different libraries of nuclear reaction rates.
Different types of QSE-groups can emerge in explosive burning.  The high
temperature phase of the rp-process (rapid proton capture) in X-ray bursts,
where no neutrons are available, witnesses isotonic lines in a 
($p,\gamma)\rightleftharpoons(\gamma,p)$ equilibrium
\cite{schatz98a,rembges97}.  Another application is the r-process,
the focus of the present paper. Here the QSE-groups are isotopic lines in an
($n,\gamma)\rightleftharpoons(\gamma,n)$ equilibrium. The connecting weak 
interactions for both processes are  $\beta^+$- or $\beta^-$-decays. During 
the final freeze-out from equilibria, when temperatures
decline below equilibrium conditions, reaction rates and capture
cross sections can count again.
In general, it should be pointed out that equilibria simplify the understanding
of explosive nucleosynthesis processes and individual cross sections play a much
less important role than reaction Q-values. Since the rp- and r-process
explore exotic nuclei close to the proton or neutron drip-lines, 
we need to focus on nuclear masses and $\beta$-decay
properties.  An understanding of the underlying nuclear structure is essential
for both types of nuclear parameters. Such information usually comes from
a close interplay between experiment and theory.
This will be discussed in more detail in sections 3 and 4.

\subsection{Necessary Environment Properties for r-Process Nucleosynthesis}

In order to underline why (within an understanding of the - nuclear -
functioning of the r-process) the two scenarios mentioned in the
beginning of this section  seem
to be the most promising ones, we want to present here two major
results found recently by Freiburghaus et al. \cite{freiburghaus99}.
They performed an analysis of neutron to seed ratios in
different types of adiabatic expansions with a given entropy S,
proton/nucleon ratio $Y_e$, and expansion time scale.
In order to obtain the required 10 to 150 neutrons per r-process
seed nucleus (in the Fe-peak or somewhat beyond) permitting to produce nuclei
with masses A$>$200, this translates into a $Y_e=$$<$$Z/A$$>$$=$0.12--0.3
for a composition of Fe-group nuclei and free neutrons.  Such a
high neutron excess is only available for high densities in neutron stars 
under $\beta$-equilibrium
($e^-+p\leftrightarrow n+\nu$, $\mu_e+\mu_p=\mu_n$),
based on the high electron Fermi energies which are
comparable to the neutron-proton mass difference \cite{meyer89}.

Another option is a so-called extremely $\alpha$-rich freeze-out in complete
Si-burning with moderate $Y_e$$>$0.40. This corresponds to a freeze-out from
QSE with a weak connection between the light ($n, p, \alpha$) and 
heavy (Mg--Kr) QSE groups due to low densities. The links accross the 
particle-unstable A=5 and 8 gaps are only possible via the three-body 
reactions $\alpha\alpha\alpha$ and $\alpha\alpha n$ to $^{12}$C and
$^9$Be, whose reaction rates show a quadratic density dependence.
The entropy can be used as a measure of the ratio between the remaining
He mass-fraction and heavy nuclei.
Similarly, the ratio of neutrons to heavy nuclei
(i.e.~the neutron to seed ratio) is a function of entropy and permits for
high entropies, with large remaining He and neutron abundances compared to
small heavy-seed abundances, neutron captures which proceed
to form the heaviest r-process 
nuclei~\cite{woosley92,takahashi94,woosley94b,hoffman96,hoffman97}.
After the freeze-out of charged-particle reactions
in matter, which expands from high temperatures but relatively low
densities, as much as 90\% of all matter can be locked into $^4$He with
$N$=$Z$, which leaves even for moderate $Y_e$'s a large neutron/seed ratio for
the few existing heavier nuclei. 

\begin{figure}
\centerline{\psfig{file=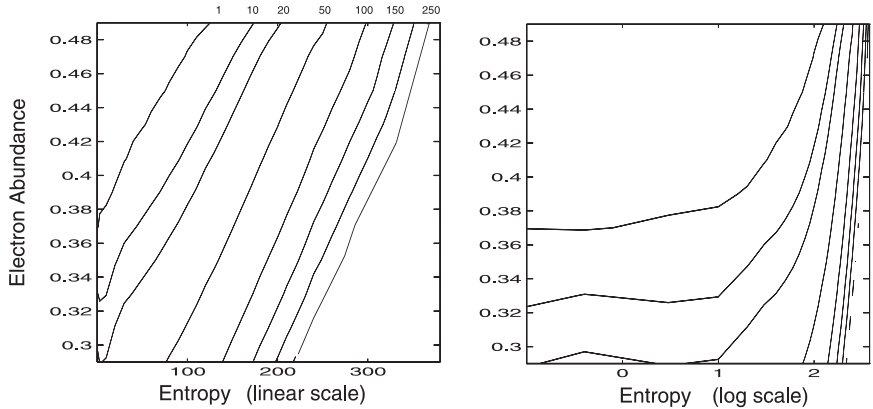,angle=0,width=12.9cm}}
\caption{$Y_n/Y_{seed}$ contour plots as a function of intial entropy
$S$ and electron abundance $Y_e$ for an expansion time scale of 0.05s.
The left part shows how, for moderate $Y_e$-values,
an increasing neutron/seed ratio - indicated by contour lines labeled with
the respective $Y_n/Y_{seed}$ -- can be attained with increasing entropy.
The results scale with
$Y_e$, measuring the global proton/nucleon ratio. The right part
of the figure enhances on a logarithmic scale the low-entropy behaviour, where
$Y_n/Y_{seed}$ is only determined by $Y_e$. The contour lines are the same for 
both figures.
\label{ynyseedplot}}
\end{figure}

The behaviour of these two environments, representing a normal (low-entropy)
and an $\alpha$-rich (high-entropy) freeze-out, is summarized in
Fig.~1. The available number of neutrons per heavy nucleus $Y_n/Y_{seed}$
after charged-particle freeze-out, when the large QSE-groups break up into
isotopic lines, is shown as a function of entropy and
initial $Y_e$. At low entropies the transition to a normal freeze-out
occurs, indicated by the negligible entropy dependence.

This introduction centered (in a cursory way) on the questions:
(i) How can one obtain - after an expansion and cooling to temperatures
of $2-3\times 10^9$K, when the charged-particle freeze-out occurs -
the neutron to seed ratios required for a successful r-process; and
(ii) what type of nuclear-physics input is of dominant importance.
In the following section 2, we will discuss in more detail the methods to
follow abundance changes during the neutron-capture phase, and will show
how in simplified r-process models the observed abundance features and nuclear
properties far from stability are related. In section 3, we present the
nuclear-data input employed, and show the strong effects
which the choice of nuclear data can have. Section 4 summarizes the experimental
information on r-process nuclei, to a large extent obtained at CERN/ISOLDE.
Finally, in section 5 we discuss how r-process calculations
evolved over the past decade with continuously improved nuclear-structure
knowledge from initially classical equilibrium calculations to
recent results from specific astrophysical sites.

\section{r-Process Nucleosynthesis Calculations}

The system of differential equations for an r-process network includes the
terms for neutron captures, neutron-induced fission, photodisintegrations, 
$\beta$-decays, $\beta$-delayed neutron emission and
fission. This leads to a set of N differential equations for the individual 
nuclear abundances as in Eq.(3)
\begin{equation}
\dot Y(Z,A)=\sum_{Z^{'},A^{'}} \lambda^{Z^{'},A^{'}} Y(Z^{'},A^{'}) +
\sum_{Z^{'},A^{'}} \rho N_A <\sigma v>_{Z^{'},A^{'}} 
Y(Z^{'},A^{'}) Y_n, 
\end{equation}
where the first term on the right-hand side includes $\beta$-decays
(with all possible emission channels, including delayed fission),
spontaneous fission, and photodisintegrations.
The second term  includes all 
neutron-induced reactions. The terms are positive or negative,
depending upon whether they produce or destroy the nucleus ($Z,A$).
For the heavy (i.e. high-$Z$) and neutron-rich nuclei in question, 
due to the high Coulomb barriers between the particles only neutron-induced
reactions are of importance for the temperatures of interest.
Thus, only reactions involving neutrons as projectiles or emitted particles,
such as ($\gamma$,n)-photodisintegrations and $\beta$-delayed neutron emission,
need to be considered. We can replace $\rho N_A Y_n$ in Eq.(3) by the neutron 
number density $n_n$, underlining the fact that $n_n$ is the important 
quantity for the r-process rather than the neutron abundance.
For nuclei with $Z$$<$80, fission does not play any role, and
neutron captures, photodisintegrations and $\beta$-decays (including
delayed-neutron branchings) dominate.
If we include $\beta$-decays leading to the emission of up to three delayed 
neutrons, the abundance change of nucleus (Z,A) is given by
\begin{subequation}
\begin{eqnarray}
\dot Y(Z,A) & = & n_nY(Z,A-1) \sigma_{A-1} + Y(Z,A+1) \lambda_{A+1}         \\
            &   & -Y(Z,A) (n_n \sigma_A + \lambda_A + \lambda_{\beta}^A +
                  \lambda_{\beta n}^A + \lambda_{\beta 2n}^A + \lambda_{\beta 
                  3n}^A)                                                   \\
            &   & +Y(Z-1,A) \lambda_{\beta}^{Z-1,A} + Y(Z-1,A+1) 
                  \lambda_{\beta n}^{Z-1,A+1}                              \\
            &   & + Y(Z-1,A+2)\lambda_{\beta 2n}^{Z-1,A+2}
                  + Y(Z-1, A+3) \lambda_{\beta 3n}^{Z-1,A+3}, 
\end{eqnarray}
\end{subequation}
where $\sigma_{A-1}$ is the thermally averaged 
$(n,\gamma)$-reaction rate $\langle \sigma v \rangle$ of nucleus ($Z,A$--1),
$\lambda_{A+1}$ is the photodisintegration rate $(\gamma, n)$ for nucleus
($Z,A$+1), $\lambda_{\beta}^A$ is the $\beta$-decay rate of nucleus $(Z,A)$, and
$\lambda_{\beta n}^A$, $\lambda_{\beta 2n}^A$, and $\lambda_{\beta 3n}^A$ 
are the rates of $\beta$-decay followed by the emission of one, two or three 
neutrons, respectively.

Depending upon the specific conditions, either $\beta$-decays can be faster
than neutron captures and photodisintegrations like in an s-process, or vice 
versa, when an
($n,\gamma$)${\rightleftharpoons}$($\gamma,n$) equilibrium exists [see sect.~1
and Eq(1)]. If the $\beta$-flow (i.e. the $\beta$-decays of the nuclei) from
each $Z$-chain to ($Z$+1) is equal to the flow from ($Z$+1) to ($Z$+2), then a
steady-flow or $\beta$-flow equilibrium will exist. When an
($n,\gamma$)${\rightleftharpoons}$($\gamma,n$) equilibrium
or a steady-flow condition is established, Eq.(4) simplifies. With these 
simplifications, a solution of the general set of equations, which would 
involve a linear system of the size of a nuclear network, can be avoided.

As long as there is a high neutron
density and a large high-energy photon density (i.e. a high temperature), we
can expect that neutron captures and photodisintegrations occur on a much 
faster time scale than $\beta$-decays. This leads after charged-particle 
freeze-out (see sect.~1) to QSE groups in form of isotopic chains (denoted
($n,\gamma$)${\rightleftharpoons}$($\gamma,n$) equilibrium or 
{\it {`waiting-point'}} approximation), where the nucleus with maximum abundance
in each isotopic chain must wait for the longer $\beta$-decay time scale.
The chemical equilibrium $\mu_n+\mu_{Z,A}=\mu_{Z,A+1}$ of particles in a
Boltzmann gas
causes abundance ratios which are dependent only on $n_n$, $T$ and $S_n$ 
\begin{equation}
{{Y(Z,A+1)}\over {Y(Z,A)}}=n_n {{G(Z,A+1)}\over{2G(Z,A)}} 
\Bigl[{{A+1}\over {A}}\Bigr]^{3/2} \Bigl[{{2\pi
\hbar^2}\over {m_ukT}}\Bigr]^{3/2} {\rm exp}(S_n(A+1)/kT).
\end{equation}
This equation is identical to Eq.(1) with $p$(projectile)=$n$,
$A_p$=1, $A_t$=$A$, $A_c$=$A+1$, and $g_n$=2. The neutron-separation energy
$S_n$ introduces the dependence on nuclear masses
in an ($n,\gamma$)${\rightleftharpoons}$($\gamma,n$) equilibrium, where
the abundance maxima in each isotopic chain are located at
the $S_n$ given by Eq.(2), which is the same in all isotopic chains.
Fig.~2 shows such a contour line of $S_n$$\simeq$2MeV for the  
Finite Range Droplet Model, FRDM \cite{moeller95}. 
In addition it displays the line of 
stability and the solar r-abundances as a function of A.

\begin{figure}
\centerline{\epsfig{file=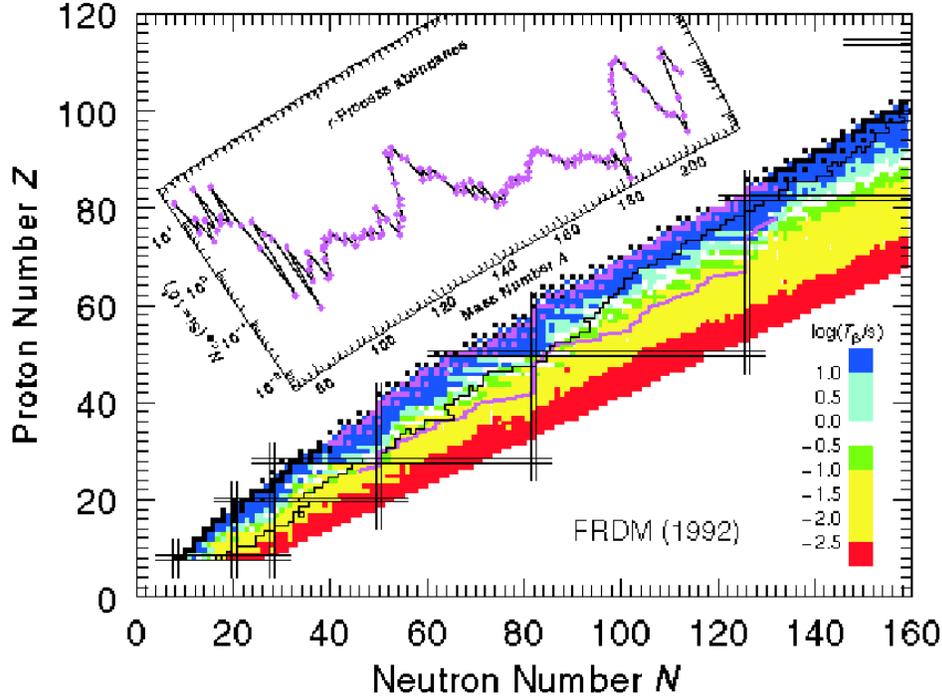,angle=0,width=12.9cm}}
\caption{Schematic illustration of the r-process path (dark line on the
neutron-rich side of $\beta$-stability) and observed r-abundances (insert).
Sharp peaks occur near A$\simeq$80, 130, and 195, where the r-process
path crosses the N=50, 82, and 126 magic neutron numbers. The chart of
neutron-rich nuclides is shaded according to measured 
and predicted $\beta$-decay half-lives, $T_{1/2}$. Grey scales for $T_{1/2}$
ranges are explained in the legend bar.}
\end{figure}

The abundance flow from each isotopic chain to the next is governed by
$\beta$-decays. We can introduce a total abundance in each isotopic chain
$Y(Z)$=$\sum_A Y(Z,A)$, and each $Y(Z,A)$ can be expressed as
$Y(Z,A)$=$P(Z,A) Y(Z)$. The individual population coefficients $P(Z,A)$
are obtained from the equilibrium condition [Eq.(5)]. The set of
differential equations which replaces Eq.(4) in this case is
\begin{equation}
\dot Y(Z)=Y(Z-1)\sum_A P(Z-1,A) \lambda_\beta^{Z-1,A} -
Y(Z)\sum_A P(Z,A) \lambda_\beta^{Z,A}, 
\end{equation}
where $\lambda_\beta^{Z,A}$ is the $\beta$-decay rate of nucleus $(Z,A)$.
This is a system of as many equations as the number of $Z$-chains. 
The individual abundances are obtained from Eq.(5). 

The waiting-point approximation is only valid for high temperatures and
neutron number densities of the gas before the freeze-out from chemical
equilibrium with neutrons occurs 
\cite{cameron83,bouquelle96})
for neutron number densities and temperatures as low as 
$n_n$$\simeq$10$^{20}$ cm$^{-3}$  and $T$$\simeq$10$^9$ K. 
Process timescales in excess of $\beta$-decay half-lives lead to
a steady-flow equilibrium, in 
addition to an ($n,\gamma$)${\rightleftharpoons}$($\gamma,n$) equilibrium,
i.e. $\dot Y(Z)$=0 in Eq.(6) or 
\begin{equation}
Y(Z)\sum_A P(Z,A) \lambda_\beta^{Z,A}= 
Y(Z)\lambda_\beta(Z)={\rm const}.
\end{equation}

In such a steady-flow equilibrium,   
the assumption of an abundance for $Y(Z_{min})$ at a minimum $Z$-value is
sufficient to predict the r-process curve. The calculation of $P(Z,A)$ from 
Eq.(5) requires the knowledge of nuclear masses (or $S_n$-values), a neutron 
number density $n_n$ and a temperature $T$. The $\beta$-decay rates in Eqs.(4),
(6) and (7) are related to the half-lives of very neutron-rich nuclei via
$\lambda_\beta=ln(2)/T_{1/2}$. Thus, in the simplest of all cases, i.e.
provided that an ($n,\gamma$)${\rightleftharpoons}$($\gamma,n$) equilibrium 
and a steady-flow has been reached, the knowledge of $n_n$, $T$, $S_n$, and
$T_{1/2}$ alone would be sufficient to predict the whole set of 
abundances as a function of $A$.  Results by Kratz et al. 
\cite{kratzea88,kratz88,kratzea93} 
indicated that this be the case for conditions which
reproduce the low-mass wings of the $A$$\simeq$80 and $A$$\simeq$130 peak 
regions of the solar-system r-process abundances $N_{r,\odot}$
\cite{kaeppeler89,beer97}.

In the most general case of (astrophysical) environment conditions, we have
thus to solve the set of Eq.(4) for all nuclei from stability to the neutron 
drip-line (1000-2000). The type of conditions after charged-particle 
freeze-out discussed in section 1.2 permits, however, the use
of Eq.(6), i.e. the assumption of an 
($n,\gamma$)${\rightleftharpoons}$($\gamma,n$) equilibrium. For small 
$\beta$-decay half-lives, encountered in between
magic numbers and for small Z's at magic numbers, even the steady-flow 
approximation of Eq.(7) seems applicable, until the final neutron and 
temperature freeze-out no longer permits the use of any equilibrium approach. 
Here, one can imagine two extreme options: (i) The occurance of an 
instantaneous freeze-out; then the results of  Eq.(6) just
have to be followed by the final decay back to stability, where
also $\beta$-delayed properties (neutron emission and fission) are needed.
(ii) In the more general case of a slow freeze-out, the set of Eq.(4) has to be
solved. Then also neutron captures can affect the initial waiting-point
abundance results.

\begin{figure}
\centerline{\epsfig{file=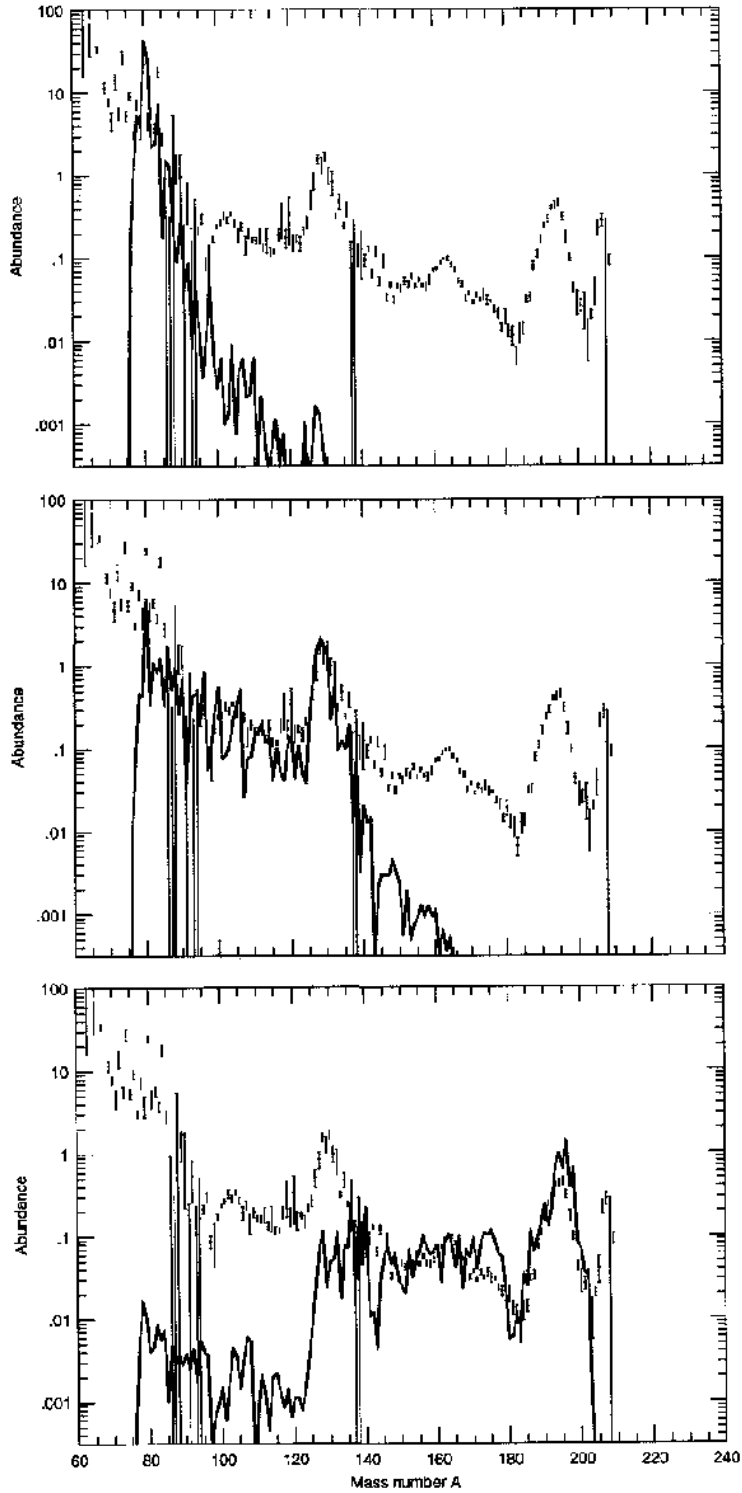,angle=0,width=10cm}}
\caption{Results of time-dependent r-process calculations with 
 $n_n$=$10^{20}$, $10^{22}$, and $10^{24}$ g cm$^{-3}$ at
$T$=1.35$\times 10^9$ K for duration times $\tau$ of 1.2, 1.7, and 2.1 s,
respectively, in comparison with solar r-process abundances 
{\protect\cite{kratzea93}}.}
\end{figure}

\section{Testing Nuclear Physics Models}

For testing nuclear-structure properties it is advisable to make use of 
a simple astrophysical environment model, with $n_n$=const, $T$=const
and a duration time $\tau$, but chosen for conditions where an equilibrium
in isotopic chains is obtained and Eq.(6) can be applied (however, without
assuming the steady flow of Eq.(7)). Fig. 3 shows three such calculations
for $n_n$=$10^{20}$, $10^{22}$, and $10^{24}$ g cm$^{-3}$ at 
$T$=1.35$\times 10^9$ K for duration times $\tau$ of 1.2, 1.7, and 2.1 s,
respectively. When testing the resulting abundances, one notices that a
steady flow [Eq.(7)] is actually obtained for the low-mass wings of the
peaks and the connecting lighter-mass regions between magic neutron numbers.
When entering these conditions, which reproduce the three r-process
peaks, into Eq.(2), one notices that they correspond to three different
r-process paths with different neutron-separation energies $S_n$ between
4 and 2 MeV. This indicates that a full fit to the solar r-process
abundances requires a superposition of different stellar conditions (but not 
necessarily different astrophysical sites).

\subsection{Nuclear Physics Input}

As outlined in the previous section, for a given neutron density (or radiation 
entropy), the r-process proceeds along a contour line of constant $S_n$ to 
heavier nuclei (see, e.g. Fig.~4 in \cite{thielemann94a}, where the abundance 
flow from each isotopic chain to the next is dominated by the $\beta$-decays. 
The flow equilibrium then implies approximate equality of progenitor abundance 
$N_{r,prog}$ (i.e. the initial abundance of a waiting-point 
isotope lying in the r-process path) times $\beta$-decay rate $\lambda_\beta$. 
Thus, with $N_{r,prog}(Z)\lambda_\beta (Z)$=const. 
the half-lives along the r-process flow path would directly define the 
$N_{r,prog}$, and -- when taking into account $P_n$ branching during 
freeze-out -- 
also the final $N_{r,\odot}$. Still some 15 years ago, there were severe doubts 
among astrophysicists  
that the validity of this simple and elegant approximation would 
ever be proven, because it requires experimental information on 
far-unstable waiting-point isotopes believed to be inaccessible in 
terrestrial laboratories. However, only a few years later in 1986, 
a new area in nuclear 
astrophysics started with the identification of the first two classical 
neutron-magic waiting-point isotopes $^{130}$Cd$_{82}$ at CERN/ISOLDE 
\cite{ref:Kratz86}, and $^{80}$Zn$_{50}$ at the fission-product mass 
separators OSIRIS \cite{ref:Lund86} and TRISTAN 
\cite{ref:Gill86}. As was shown by Kratz et al. 
\cite{ref:Kratz86,kratzea88,kratz88}, indeed, first evidence for the existence 
of -- at least a local -- steady-flow equilibrium could be achieved, which 
presented a stimulating challenge to both theoreticians and 
experimentalists in the nuclear-physics community in the following years.

Nevertheless, even today the vast majority of very neutron-rich r-process 
nuclei is experimentally not accessible. Therefore, 
a general understanding of 
their nuclear properties remains to be obtained only through theoretical 
means. Since a number of different quantities are needed in 
r-process calculations, as outlined above, in the past it was often not 
possible to derive them all from one source. Taking them from different 
sources, however, may raise the question of consistency. In order to avoid 
an eventual vanishing of real signatures or the creation of artificial 
r-abundance effects from the use of nuclear-physics parameters from 
different models -- sometimes even of largely different sophistication --
we have performed our r-process calculations in a unified 
macroscopic-microscopic approach 
in which all nuclear properties can be studied in an internally 
consistent way. This approach is, for example, discussed in detail in 
Kratz et al. \cite{kratzea93} and M\"oller et al. \cite{ref:MNKz97} for the 
combination of nuclear masses from the Finite Range Droplet Model, FRDM 
and $\beta$-decay properties from a Quasi-Particle Random Phase Approximation, 
QRPA \cite{krumlinde}, in its present version using Folded-Yukawa 
single-particle levels and the Lipkin-Nogami pairing model 
(M\"oller and Randrup \cite{moran90}). Analoguously, when 
adopting other mass formulae, such as the Extended Thomas-Fermi plus 
Strutinsky Integral models, ETFSI-1 (Aboussir et al. \cite{aboussir95}) or 
ETFSI-Q (Pearson et al. \cite{pearson96}), or the spherical  
Hartree-Fock-Bogolyubov method with a specific Skyrme force, HFB/SkP 
(Dobaczewski et al. \cite{dobaczewski94,dobaczewski95}), we 
normally use theoretical $\beta$-decay quantities deduced from QRPA 
calculations with masses and deformation parameters given by that 
particular model. 

With the above macroscopic-microscopic approaches, several rather 
sophisticated and internally consistent, global nuclear-data sets are now 
available for astrophysical calculations which are expected to yield 
more reliable predictions than earlier models, like the old droplet-type 
formulae of Myers \cite{myers76} and Von Groote et al. \cite{groote76} or 
the Gross Theory of $\beta$-decay (Takahashi et al. \cite{takahashi73}). 
Nevertheless, being aware that even the more microscopic approaches do have 
their deficiencies, we have continuously tried to improve the basic data sets 
by short-range extrapolations of known nuclear-structure properties, 
either model-inherently not contained in or not properly described by 
the above global methods. For details, see e.g. Kratz et al. 
\cite{kratzea93}. And, in contrast to other authors who prefer to use 
exclusively model predictions in their r-process calculations (see, e.g. 
\cite{bouquelle96}), we have steadily updated our 
files by including all new experimental masses, $T_{1/2}$ and $P_n$ values 
(\cite{ref:Audi97,klkr96}).

These nuclear-data sets have been used extensively during the last years 
in our r-process calculations 
(\cite{kratzea93,thielemann94a,chen95,ref:Kratz98,pfeiffer97,freiburghaus97,kratz98,freiburghaus99}) 
to reproduce the $N_{r,\odot}$ pattern \cite{kaeppeler89,beer97} and to test 
nuclear structure close to the neutron drip-line.  

\subsection{Testing Nuclear Physics Models}

As has been outlined in sect.~2, calculated r-abundances are strongly 
dependent on model predictions of masses and $\beta$-decay properties. 
Therefore, it seems worthwhile to test the influence of the choice of 
nuclear theories in selected mass regions before performing global, 
multi-component r-process calculations. Such tests may well be able to 
shed some light on the development of nuclear structures towards the 
neutron drip-line, and with this may define certain key properties and 
key experiments. Of particular importance in this context are the 
N$_{r,\odot}$ peak areas, since the related N=50, 82 and 126 shell 
closures act as 'bottle necks' for the r-process matter flow.

Already Burbidge et al. \cite{b2fh} pointed out that the r-process very 
likely passes through neutron-magic isotopes around A$\simeq$80, 130 and 
195 which have longer-than-average $\beta$-decay half-lives. At these points 
the r-abundances build up and form the observed N$_{r,\odot}$ peaks, as 
nuclei like $^{80}$Zn$_{50}$, $^{130}$Cd$_{82}$ and $^{195}$Tm$_{126}$ 
'wait' to $\beta$-decay. It is therefore immediately evident that these 
clasical 'waiting points' would not only give important constraints on 
r-process conditions, but would also yield much needed information on shell 
struture far from $\beta$-stability.

In fact both, the 'long' T$_{1/2}$ of the above three classical waiting-point 
nuclei, as well as the 'low' S$_n$ values of the respective (N$_{mag}$+1) 
isotopes will considerably slow down the nuclear flow through the A$\simeq$80, 
130 and 195 mass regions, and with this will to a large extent determine the 
total duration of a 'canonical' r-process \cite{b2fh,mathward85,kratz88}. To 
first illustrate this for the $\beta$-decay half-lives, Fig.~4 shows
four calculations of an r-process component
mainly responsible for the build-up of r-abundances in the 
A$\simeq$130 peak region, where all 
stellar (T$_9$, n$_n$, $\tau$) and nuclear-physics (Q$_{\beta}$, S$_n$, 
$\epsilon_2$, P$_n$) input parameters (taken from 
\cite{moeller95,ref:Audi97,klkr96}), 
except the $\beta$-decay 
half-lives of $^{129}$Ag$_{82}$ and $^{130}$Cd$_{82}$, are kept 
constant. The T$_{1/2}$ of these two waiting-point isotopes are varied 
within the range of earlier model predictions. It is quite evident from 
this figure, that the T$_{1/2}$ not only determine the height (and shape) 
of the A$\simeq$130 N$_{r,\odot}$ peak, but also the r-matter flow to higher 
masses. With `short' half-lives (T$_{1/2}$=20~ms; see upper 
left part of Fig.~4), 
the r-process would pass the N=82 region more or less unhindered; this 
would result in a 'short' overall process duration. On the other hand, 
`long' half-lives (T$_{1/2}$=750~ms; see lower right part of 
Fig.~4) would slow down the 
nuclear flow through the A${\simeq}$130 region considerably, and 
consequently would require an r-process duration of 5--10~s when starting 
from Z=26. Such time scales are, 
however, too long for the presently favoured r-process scenarios. 
With this, also the reliability of gross $\beta$-decay theories which 
yield model-inherently `longer' T$_{1/2}$ far from stability 
\cite{takahashi73,tach90}, may be questioned.

\begin{figure}
\centerline{\epsfig{file=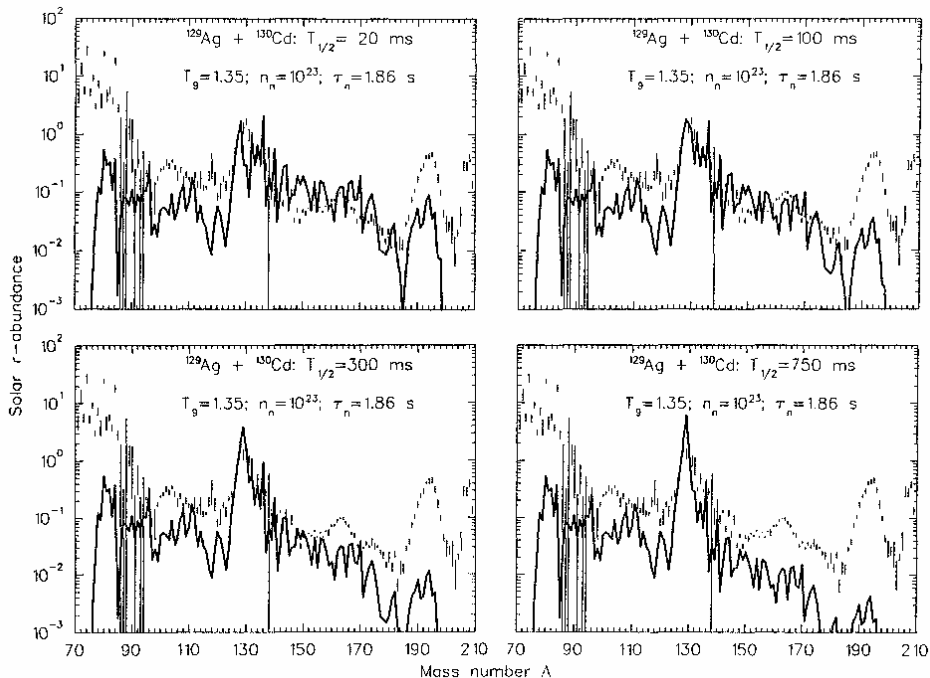,angle=0,width=12.9cm}}
\caption{
Time-dependent calculations of an r-process component which is mainly 
responsible for the build-up of abundances in the A$\simeq$130 region. All 
stellar and nuclear-physics input parameters, except the $\beta$-decay 
half-lives of the two N=82 waiting-point isotopes $^{129}$Ag and $^{130}$Cd, 
are kept constant. For details, see text. 
}
\end{figure}

We now want to test the behaviour of different mass models; i.e.~the 
old Hilf~et~al.~formula \cite{hilf76}, the two global 
macroscopic-microscopic theories FRDM \cite{moeller95} and ETFSI-1 
\cite{aboussir95}, and the recent microscopic spherical HFB+SkP model 
\cite{dob96}. Within a very simple {\it static} approximation (see, 
e.g.~\cite{kratzea88,kratzea93}), one can obtain a first rough estimate under 
which neutron-density 
conditions an r-process will start to break through the classical 
waiting points at N=50, 82 and 126. As can be seen from Fig.~5, 
depending on the magic shell there are considerable differences for the 
four mass models chosen. The behaviour is clearly correlated with the 
respective shell strengths. A strong shell closure is reflected by a low 
neutron separation energy (in Fig.~5 given as 
$\overline{S}$$_n$=[S$_n$(A)+S$_n$(A+1)]/2) and a high neutron density 
required to overcome the neutron-capture `bottle-necks' at 
$^{80}$Zn, $^{130}$Cd and $^{195}$Tm. When focussing for example on N=82, 
the strongest shell closure is predicted by the ETFSI-1 model, whereas 
the recent HFB+SkP indicates a much weaker magicity. Already from 
this simple picture it is evident, that for the production of the whole 
N$_{r,\odot}$ distribution quite different neutron-density regimes 
will be required for different mass models.

\begin{figure}
\centerline{\epsfig{file=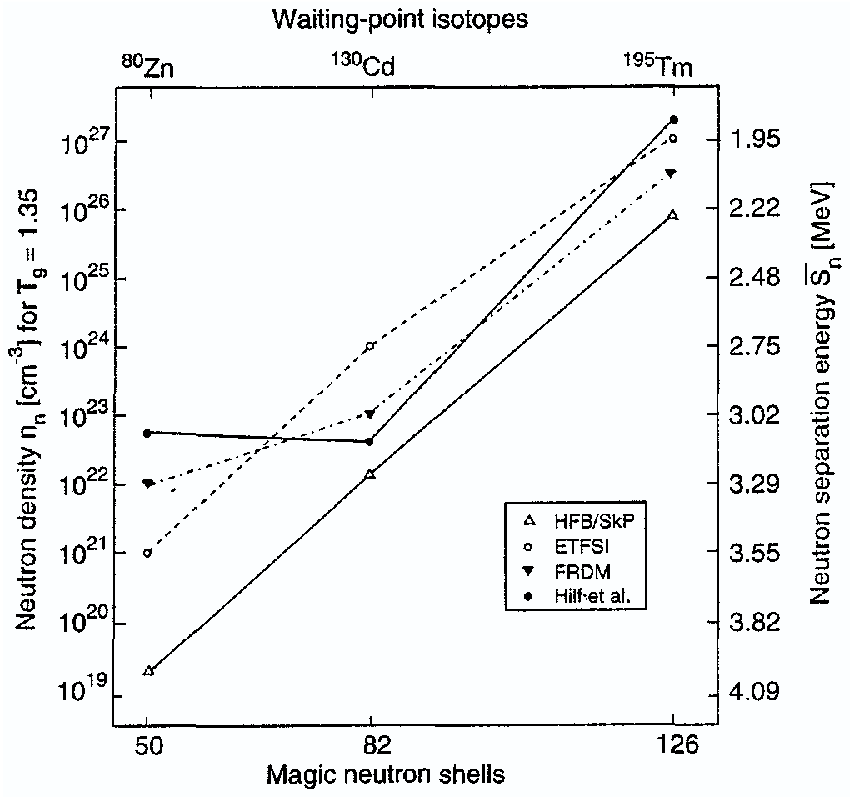,angle=0,width=8.5cm}}
\caption{
Static equilibrium calculations of n$_n$ conditions 
for the break-through of an r-process at the classical waiting-point 
isotopes $^{80}$Zn$_{50}$, $^{130}$Cd$_{82}$ and $^{195}$Tm$_{126}$ for a 
stellar temperature of T$_9$=1.35. The four mass models chosen 
{\protect \cite{hilf76,moeller95,aboussir95,dob96}} show a different 
behaviour. Depending on the strength of the shell 
closure, the break-through occurs at lower n$_n$ (e.g.~at N=50, already 
around n$_n$$\simeq$10$^{19}$~cm$^{-3}$ for the ''quenched'' HFB/SkP 
masses) or at 
higher n$_n$ (again at N=50, only at n$_n$$\simeq$10$^{23}$~cm$^{-3}$ for 
the Hilf~et~al.~masses). The differences in n$_n$ are correlated 
with different r-process paths, as indicated by the different average 
$\overline{S}$$_n$ values at the right-hand axis of ordinates. 
}
\end{figure}

In a next step, we will check the above indication by more realistic 
time-dependent calculations. In Fig.~6, the results of 
four one-component r-process calculations are shown, which use exactly 
the same astrophysical parameters as 
in our T$_{1/2}$ test to produce the A$\simeq$130 r-abundance region 
(see Fig.~4). In all calculations, the $\beta$-decay 
properties were now taken from our recent evaluation \cite{klkr96}. In this 
picture, the decisive nuclear-physics quantity are the S$_n$ values, 
in particular those  of 
the  waiting-point isotopes before and at N=82. Apart from a certain 
memory effect from the N=50 shell closure, it is quite obvious  
that the predicted N=82 shell strength (see Fig.~5) determines the 
r-matter flow in the A$\simeq$130 peak region and beyond. For mass 
models with a pronounced N=82 shell closure, i.e.~FRDM and ETFSI-1, a 
significant r-abundance trough occurs around A$\simeq$115, and the low 
S$_n$'s of the N=83 isotones of $_{41}$Nb to $_{48}$Cd prevent a rapid 
matter flow to heavier masses under the given astrophysical conditions. 
\begin{figure}
\centerline{\epsfig{file=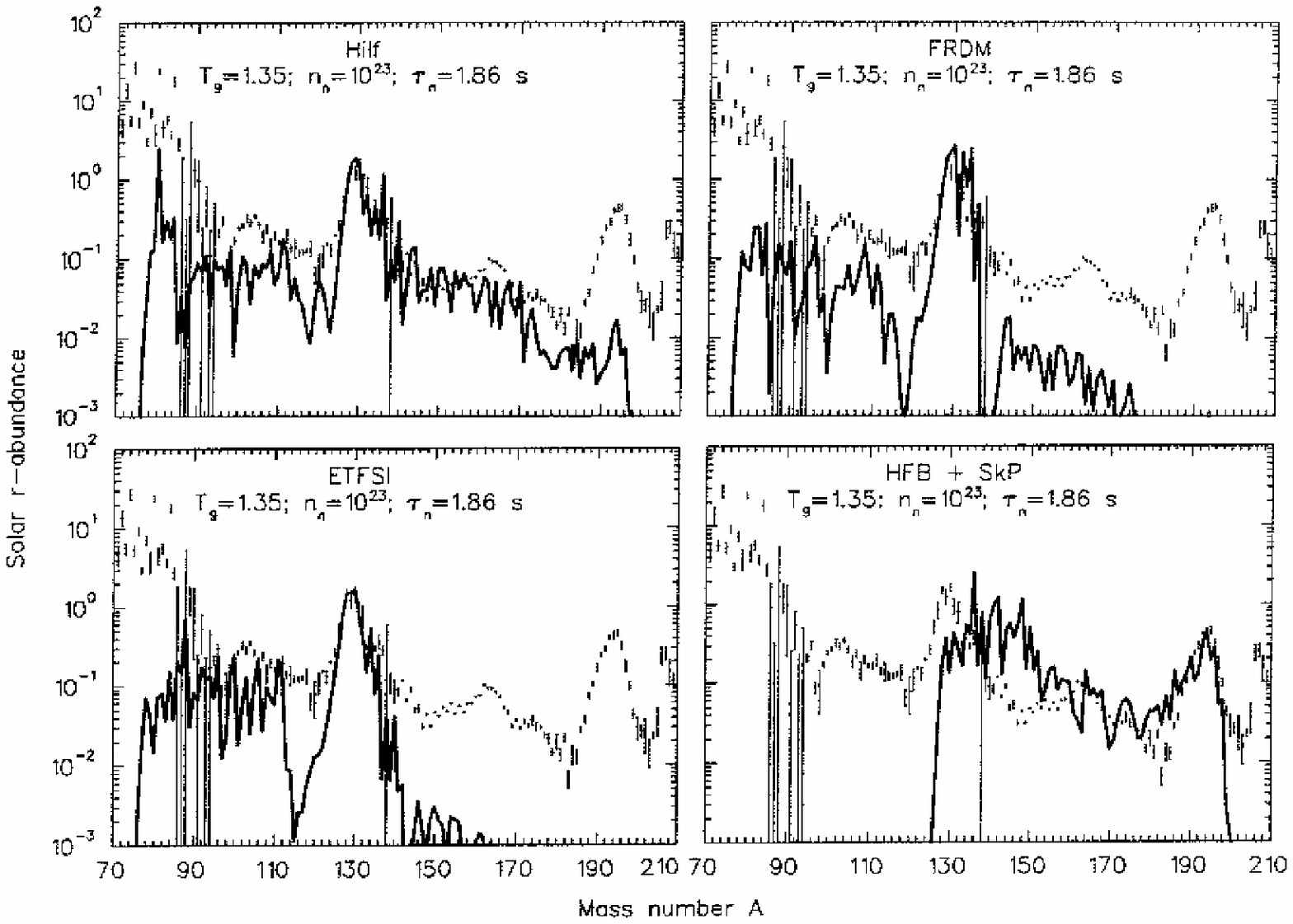,angle=0,width=12.9cm}}
\caption{
Time-dependent calculations of an r-process component with S$_n$ values from 
four different mass models 
{\protect \cite{hilf76,moeller95,aboussir95,dobaczewski95}}. The 
astrophysical parameters T$_9$, n$_n$ and $\tau$ are the same as in Fig.~4; 
all nuclear-physics data are taken from {\protect \cite{klkr96}}. In this 
picture, the nuclear flow into the 
A$\simeq$130 abundance peak and the mass region beyond is determined by 
the strength of the N=82 shell, reflected here by the S$_n$ values of 
the N=81 and 83 isotones of $_{41}$Nb to $_{48}$Cd (see, e.g. Fig.~4 in
{\protect\cite{chen95}}). For further discussion, see text. }
\end{figure}
On the other hand, a weaker N=82 shell closure, as e.g.~predicted by the 
Hilf et al.~formula and the recent HFB/SkP theory, allows a more rapid 
build-up and transit of the A$\simeq$130 region. With these mass models, 
the N=82 magic shell is reached at a higher Z (i.e.~not before $_{43}$Tc), 
thus avoiding the A$\simeq$115 abundance deficiency. And, at the same 
time, due to their somewhat higher S$_n$(N=83) values the break-out 
from N=82 can occur already at a lower neutron density (i.e.~around 
n$_n$$\simeq$1--5$\times$10$^{22}$~cm$^{-3}$, according to Fig.~5). We see from 
Fig.~6, that in the `extreme' of the HFB/SkP theory with 
Bogolyubov-enhanced shell-quenching towards the neutron drip-line, 
a neutron density of n$_n$$\simeq$10$^{23}$~cm$^{-3}$ and a time scale 
of $\tau$$\simeq$1.9~s are already sufficient to drive the nuclear flow from Fe 
up to the A$\simeq$195 r-abundance peak, whereas with the FRDM and 
ETFSI-1 mass models just the A$\simeq$130 peak is built up. This result 
may be of particular interest for the recently favoured high-entropy 
bubble scenario of a type-II supernova \cite{woosley92,meye92,takahashi94}, 
which still encounters some problems in obtaining high enough neutron 
densities (respectively radiation entropies) to completely build up the 
A$\simeq$195 r-abundance peak. 

\section{Experimental information on r-process nuclei}

The experimental study of neutron-rich nuclides lying in and near the 
projected r-process path serves two purposes, provision 
of direct data for use in nucleosynthesis calculations, and testing the 
theories from which nuclear properties of far-unstable isotopes are 
derived when no data are available. As is infered from the   
nucleosynthesis snapshots shown  in the previous section, the 
predictions of the presently existing global nuclear models 
quite obviously differ considerably
when approaching the limits of particle binding. The reason may well 
be that the model parameters used so far, which were mainly determined 
to reproduce known properties near $\beta$-stability, need not always be 
proper to be used at the drip-lines. Therefore, experiments very far from 
stability will be essential to verify possible nuclear-structure changes 
with isospin, and to motivate improvements in  nuclear theories.

As already mentioned in sect.~3, only about a decade ago the area of 
experiments {\it in the r-process path} started with the identification 
of the first two classical, neutron-magic `waiting-point' isotopes: 
$^{80}$Zn$_{50}$ (situated 10 mass-units away from stability) and  
$^{130}$Cd$_{82}$ (even 16 units beyond stable $^{114}$Cd) 
\cite{ref:Lund86,ref:Gill86,ref:Kratz86}.

In the following, we will give some technical details on the identification 
of $^{130}$Cd, in order to convince also outsiders from our particular 
field that such, in principle quite simple experiments are not at all 'easy'. 
Neutron-rich Cd isotopes were produced at that time at the old CERN/ISOLDE 
mass separator connected to the 600-MeV proton Synchro-Cyclotron.
In this experiment, a UC$_2$ target in a 
graphite-cloth matrix was connected to a plasma-discharge ion source via a 
heated quartz tansfer line. This tube was assumed to serve as a kind of 
thermochromatography column allowing preferential extraction of volatile 
species, whereas non-volatile elements should be retained. In this way, some 
chemical selectivity was intruduced to the otherwise non-selective plasma 
ionization. Nevertheless, at A=130 strong isobaric contaminations of 
$^{130}$In and $^{130}$Cs were observed which made ${\beta}$- and  
$\gamma$-spectroscopy of the orders of magnitude weaker produced $^{130}$Cd 
impossible. Moreover, even with the selective detection method of     
$\beta$-delayed neutron ($\beta$dn) counting,
the occurrence of -- a priori unexpected -- [$^{40}$CaBr]$^+$ molecular ions, 
at A=130 containing the 1.9-s $^{90}$Br $\beta$dn-precursor, severely 
complicated the experimental conditions. To optimize the detection conditions 
for the searched $^{130}$Cd, 
data from about 36,000 growth (300 ms collections) and decay (900 ms) 
cycles were accumulated. A  careful analysis of the  resulting 
complex $\beta$dn-multiscaling curves \cite{ref:Kratz86,kratz88} yielded -- 
after subtraction of the $^{90}$Br $\beta$dn-component --
an estimate of the $^{130}$Cd half-life between maximum 230 ms 
(presumably representing a T$_{1/2}$ mixture of Cd-mother and In-daughter 
decays) and 160 ms 
(derived from the growth curve during activity collection, where the Cd 
signals should be enhanced compared to the In-daughter activity).

After all these experimental difficulties, it was quite satisfying to 
see that our experimental half-life  
value of (195$\pm$35) ms turned out to be in reasonable agreement with the 
N$_{r,\odot}$(Z)$\times$$\lambda$$_{\beta}$(Z)$\approx$const. waiting-point 
expectation of (180$\pm$20) ms, derived independently at that time by 
Hillebrandt (see, Kratz, Thielemann, Hillebrandt et al. \cite{kratzea88}). 
The correlation found between the $^{131,133}$In and $^{130}$Cd half-lives 
and the observed solar abundances of their stable isobars, which was for the 
first time exclusively based on experimental numbers, immediately 
became very important to constrain the equilibrium conditions of an 
r-process. And this success strongly motivated further experimental and 
theoretical nuclear-structure investigations, as well as astrophysical 
r-process studies \cite{kratzea93,thielemann94a}. For example, at ISOLDE the 
second N=50 waiting-point isotope $^{79}$Cu could be identified -- again using 
a chemically non-selective plasma-ion source \cite{kratz91}; 
experimentally known strong  P$_n$ branches were shown to be the 
nuclear-structure origin of the odd-even staggering in the A$\simeq$80 
N$_{r,\odot}$ peak \cite{kratz90}; and from the interpretation of the 
$^{80}$Zn decay scheme in terms of J$^{\pi}$=1$^+$, two-quasi-particle (2QP) 
configurations, evidence for a vanishing of the spherical N=50 shell closure 
far from stability was obtained \cite{ref:krat88}.

Since the late 1980's, considerable progress has been achieved in the study 
of neutron-rich medium- to heavy-mass nuclei at various laboraties, with 
ISOLDE always playing a leading role in this field. However, due to the 
generally 
very low production yields, the occurrance of isobar, multi-charged ion and 
molecular-ion contaminations from chemically non-selective ionization modes,
or because of the application of non-selective detection methods, {\it direct} 
information on isotopes lying in the r-process path(s) could be obtained but 
in a few exceptional cases. From the majority of investigations, 
originally dedicated to study nuclear-structure developments as a 
function of isospin, only {\it indirect} -- but nevertheless also 
important -- information for r-process calculations can be deduced (see, e.g. 
\cite{kratz88,kratzea93,thielemann94a,chen95,ref:Kratz98,pfeiffer97}). 
In this context, we would like to remind the reader of the 
fascinating phenomenon of the {\it sudden onset and saturation of 
ground-state deformation at N=60} (see, e.g. 
\cite{ref:Zr88,ref:IOP92,ref:ENAM95}), which had in fact led to the first 
'astrophysical request' of N=82 shell quenching \cite{kratzea93}.   

Because of the severe experimental problems to identify r-process nuclei, 
recent progress has undoubtedly benefitted from the {\bf selectivity} in
their production and detection, by applying e.g. Z-selective laser 
ion-source (LIS) systems, isobar separation or multifold-coincidence 
techniques. 

Today, there are mainly three mass regions, where nuclear-structure information 
is of particular astrophysical interest. The first is the 
seed region of the classical r-process which involves very 
neutron-rich Fe-group isotopes up to the double-magic nucleus 
$^{78}$Ni. We will discuss recent studies in that mass-region in subsect. 
4.1.  The second region of interest is that of far-unstable isotopes of 
refractory elements (Zr to Pd) around A$\simeq$115. Here, most r-process 
calculations show a pronounced r-abundance {\it trough}, which is believed 
to be due to deficiencies at and beyond N=72 mid-shell (see, e.g. 
\cite{kratzea93,chen95,pfeiffer97}. Recent 
experimental information using chemical separation procedures, or ion-guide 
and projectile-fragmentation techniques combined with mass separation can, 
for example be found in \cite{ref:ENAM95,schoed95,mehr96,lher98,ENAM98}. 
The third region of strong nuclear-structure and astrophysics 
interest is that around the double-magic nucleus $^{132}$Sn, which we will 
discussed in subsect. 4.2. 

\subsection{The $^{68}$Ni region} 
 
Recent spectroscopic studies around the 
double (semi-) magic $^{68}$Ni (Z=28, N=40) at LISOL in 
Louvain-la-Neuve, at GANIL/LISE in Caen and also at CERN/ISOLDE (see, e.g. 
\cite{franch,sorlin,musec,hannaw}, and references therein), have revealed new 
structure features which are not easy to reconcile with established shell-model 
calculations in the neutron-rich region between $^{48}$Ca and $^{78}$Ni. 
Most (but not all) of the models find that N=40 is a good closed subshell 
and predict near-zero deformation for neutron numbers in its vicinity. 
Of course, these ideas are supported by existing 
data for $_{28}$Ni, $_{30}$Zn, $_{32}$Ge and $_{34}$Se isotopes for which 
there is a slight peak in the 2$^+$ energies for N=38 and~/~or the 
supposed subshell at N=40, and 
then a gentle drop in E(2$^+$) beyond N=42 as the $\nu$g$_{9/2}$ orbitals 
fill and some collectivity is observed (see Fig.~7). 

However, recent experiments at the PS-Booster ISOLDE using the high 
Z-selectivity of laser ionization have made possible decay-studies of 
very neutron-rich $_{25}$Mn isotopes, and the determination of detailed 
nuclear-structure information for their $_{26}$Fe-daughters. In particular, 
it was found that -- contrary to expectations -- the N=40 isotone $^{66}$Fe, 
situated only one proton-pair below double (semi-) magic $^{68}$Ni, is 
already relatively deformed. From the well-known relationship between E(2$^+$),
B(E2) values and collectivity \cite{raman91}, we deduce a deformation  
value of $\beta$$_2$$\simeq$0.26 for $^{66}$Fe. 
As can be seen from Fig.~7, this trend toward collectivity below 
$^{68}$Ni is already observed for $^{64}$Fe$_{38}$ with 
$\beta$$_2$$\simeq$0.18. Needless to say, these data serve as an indication 
that `theoretical 
consensus' for sphericity in N=40 isotones with Z$<$28 does not provide a good 
description of the structure, and consequently also of nuclear masses and 
decay properties of these nuclides. Subsequent data from the GANIL/LISE 
spectrometer \cite{sorlENAM,azaiez}
in this mass region are in support of our conclusion that the N=40 and 
adjacent nuclides below $^{68}$Ni are relatively deformed, with the maximum 
collectivity presumably centered around $^{64}$Cr$_{40}$. Hence, the use 
of theoretical results that do not account for deformation of neutron-rich 
isotopes in this mass region as input data for astrophysical 
calculations must be approached with caution.

\begin{figure}
\centerline{\epsfig{file=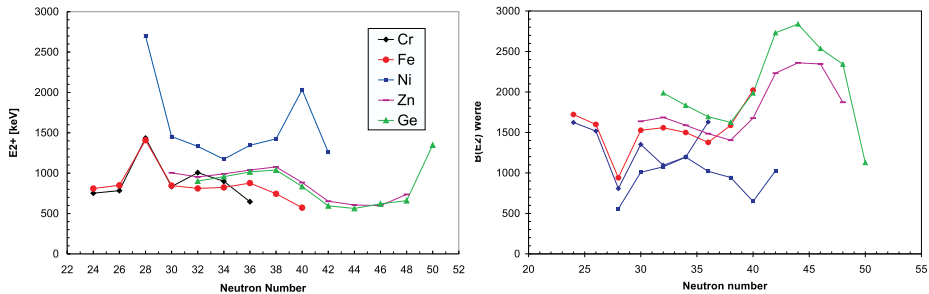,angle=0,width=12.9cm}}
\caption{
Energies of the first 2$^+$ levels and B(E2) values of 
even-even $_{24}$Cr to $_{32}$Ge nuclides.
}
\end{figure}

Already with the use of Z-selective laser ionization at the General-Purpose 
Separator (GPS) of CERN/ISOLDE, it has been possible to study decay  
properties of 14-ms $^{69}$Mn, which is one of the most 
neutron-rich isotopes with the shortest $\beta$-decay half-life  
beyond the sd-shell. The prospects are good for an extension of such 
measurements to even heavier isotopes with LIS and the High-Resolution 
Separator (HRS), eventually including the study of detailed level structures 
up to $^{70}$Fe. Combined with ongoing work on neutron-rich $_{29}$Cu and 
$_{30}$Zn at ISOLDE and on $_{27}$Co and $_{28}$Ni at other facilities, 
nuclear-structure information towards the next major neutron-shell closure 
N=50 will accumulate, to answer e.g. the question about shell-quenching at
double-magic $^{78}$Ni and its `bottle-neck' behaviour for the r-process 
matter flow through N=50. Moreover, the nuclear properties in this region have
gained additional importance in the context of the lately discussion of 
{\it two} distinct r-process components, a `weak' r-process below a$\simeq$130,
and the `main' r-process beyond A$\simeq$130 (see sect. 5.2).

\subsection{The $^{132}$Sn region}
                                                               
The mass region around the far-unstable double-magic nucleus $^{132}$Sn has 
long been and still is of particular interest for experimental and theoretical 
investigations. Apart from astrophysical importance (formation of 
the A$\simeq$130 peak of the  
N$_{r,\odot}$ distribution  \cite{kratzea93,ref:Kratz98}), this 
region is of considerable shell-structure interest. The neutron-rich isotope 
$^{132}_{\phantom{1}{50}}$Sn$_{82}$ with its pronounced magicity (only 
comparable to stable $^{208}_{\phantom{1}{82}}$Pb$_{126}$), together with 
the properties of the 
nearest-neighbour single-particle ($^{133}_{\phantom{1}{\bf 51}}$Sb$_{82}$ 
and $^{133}_{\phantom{1}{50}}$Sn$_{\bf 83}$) and single-hole 
($^{131}_{\phantom{1}{50}}$Sn$_{\bf 81}$ 
and $^{131}_{\phantom{1}{\bf 49}}$In$_{82}$) nuclides are essential for tests 
of the shell model, and as input for any reliable future microscopic 
nuclear-structure calculations towards the neutron drip-line.

The bulk of the data so far known in this region have been obtained from 
$\beta$-decay spectroscopy at the mass-separator facilities OSIRIS
(Sweden) and ISOLDE (see, e.g.
\cite{ref:Hoff96,ref:SanV98,ref:Fog98,ref:Stone97,ref:Zha97,ref:Fedo95,ref:Kaut96,ref:Kratz98a},
and references therein).
The structures of $^{131}$Sn ($\nu$-hole) and $^{133}$Sb ($\pi$-particle) 
are fairly well known since more than a decade. More recently, 
the $\nu$-particle states 2f$_{7/2}$ (g.s.), 3p$_{3/2}$ (854 keV), 
1h$_{9/2}$ (1561 keV) and 2f$_{5/2}$ (2005 keV), as well as tentatively 
also the 3p$_{1/2}$-particle (1655 keV) and the 1h$_{11/2}$-hole 
($\approx$3700 keV) states in $^{133}$Sn$_{83}$ have been identified 
\cite{ref:Hoff96} at the General Purpose Separator (GPS) of the new PS-Booster 
ISOLDE facility. From these data, valuable information 
on the spin-orbit splitting of the 2f-orbital and tentatively the 
3p-state splitting was obtained. These results were compared   
to mean-field and HFB predictions, and it was found that none of the 
potentials currently used in {\it ab initio} shell-structure calculations 
was capable of properly reproducing the ordering and spacing of these 
states (see, e.g. \cite{ref:Rau98}, where also possible astrophysical 
consequences are given). Of particular interest in this context are the 
surprisingly low-lying $\nu$p$_{3/2}$- and $\nu$p$_{1/2}$-states in 
$^{133}$Sn. According to the standard Nilsson model 
\cite{ref:RaSh84}, for example, they are expected at 2.89 MeV and 
4.36 MeV, respectively. Such lowering of the energies of low-j orbitals 
(here, by as much as 2.0 MeV and 2.7 MeV, respectively)
seems to occur in different mass regions for very neutron-rich nuclides, 
and has been interpreted, for example, as monopole shifts of 
single-particle (SP) states\cite{walters98}. The occurrence of such energy 
shifts of SP levels has recently also
been predicted as a {\it neutron-skin} phenomenon to occur 
only near the neutron drip-line (see, e.g. \cite {dobaczewski94}), but 
not yet in (N$_{mag}$+1) $^{133}$Sn which is 
still neutron-bound by S$_n$$\simeq$2.4 MeV. Nevertheless, following the 
suggestion of Dobaczewski \cite{dobaczewski94}, we have modified the Nilsson 
potential by reducing the strength of the ${\it l}^2$-term (i.e. the 
spin-orbit interaction), in order to 
study its effect on different orbitals. And indeed, this procedure has led 
to the desired change in the position and 
even the ordering of the $\nu$-particle states beyond N=82, and has 
thus allowed at least  a qualitative 
reproduction of the experimental observation of low-lying low-j and high-lying 
high-j orbitals in $^{133}$Sn. 

Recently, Zhang et al. \cite{zhang98} have proposed a new set of Nilsson 
parameters that correctly reproduce the SP levels in $^{131}$Sn and $^{133}$Sn.
They compare their new results with the experimental levels and older Nilsson
parameters, as well as SP levels obtained from relativistic mean-field
calculations. The danger with this approach is that the change in parameters
may be compensating for underlying deficiencies in the model. Consequently,
while the results do fit the levels of $^{133}$Sn, they may not be useful for
astrophysical uses where the goal is to determine the structure of nuclides
significantly further from stability.

With these new data, the $^{132}$Sn valence-nucleon region is nearly complete.
The only missing information are the $\pi$-hole states 
in $^{131}_{\phantom{1}{\bf 49}}$In, which can be studied 
through $\beta$-decay 
of the very exotic nucleus $^{131}$Cd$_{83}$. Recent LIS developments at 
Mainz \cite{ref:Erd98} and CERN/ISOLDE, using a novel frequency-tripling 
technique for the Z-selective resonance ionization of Cd, have  
made possible first test measurements on Cd isotopes in the A$\simeq$130 region
at the GPS of the PS-Booster ISOLDE. Although the data are not yet fully 
analyzed, we can already present some interesting new results. For example, 
the complex $\beta$dn-curve at A=130 indicates a half-life component of
(162$\pm$7)~ms on top of the $^{130}$In $\beta$dn-activity. We assign this 
component to the r-process waiting-point isotope $^{130}$Cd, which we had first 
identified -- using non-selective ionization -- with a somewhat longer 
half-life \cite{ref:Kratz86}. As weighted average of the two 
determinations, we now favour a T$_{1/2}$=(168$\pm$12)~ms. 

In this experiment, it was also possible to measure, for the first time, a
half-life for $^{131}$Cd$_{83}$. The value of (68$\pm$3)~ms \cite{hann} is
considerably shorter, and the estimated $\beta$dn-branch is much lower than 
might have been expected. It is possible to estimate a forbidden g.s.-to-g.s.
$\beta$-branch ($\nu$f$_{7/2}$ $\rightarrow$ $\pi$g$_{9/2}$) with a partial
half-life of about 120 ms. This means that the Gamow-Teller (GT) allowed decay
must also have a half-life of at least 150 ms, as compared to the earlier
QRPA prediction of 943 ms \cite{ref:MNKz97}. As it is unlikely that the log(ft)
value is smaller, this result indcates a significantly larger $\beta$-energy
driving the GT decay than predicted by the QRPA. In turn, this means either a 
significantly larger Q$_\beta$ (meaning a less bound 83$^{rd}$ neutron) or a
significantly lower energy for the daughter states to be populated in 
$^{131}$In.
Either of these possibilities is an indication that full understanding of the 
structure of nuclides along the N=82 closed neutron shell is not complete.

In any case, the present results from both detection 
methods, $\beta$dn-counting and $\gamma$-spectroscopy, suggest the need of 
isobar separation, i.e. the use of the HRS at ISOLDE, in order to discriminate 
the laser-ionized $_{48}$Cd from the unavoidable surface-ionized $_{49}$In 
isobar. 
 
Another recent study  of nuclear-structure development towards N=82 
concerns the $\beta$dn- and $\gamma$-spectroscopic measurements of 
neutron-rich Ag nuclides at the PS-Booster ISOLDE facility 
(\cite{ref:Zha97,ref:Fedo95,ref:Kaut96,ref:Kratz98a,KRAENAM98}), using an 
improved version of the LIS system described in \cite{ref:Mish93} together 
with new microgating procedures \cite{ref:Jad97}.
This approach was of considerable assistance in minimizing the activities 
from surface-ionized In and Cs isobars. 
In this context, the additional `selectivity' of the spin- and 
moment-dependent hyperfine (HF) splitting was used to enhance the ionization of 
either the $\pi$p$_{1/2}$-isomer or the $\pi$g$_{9/2}$ g.s.-decay of the 
Ag isotopes.

\begin{figure}
\centerline{\epsfig{file=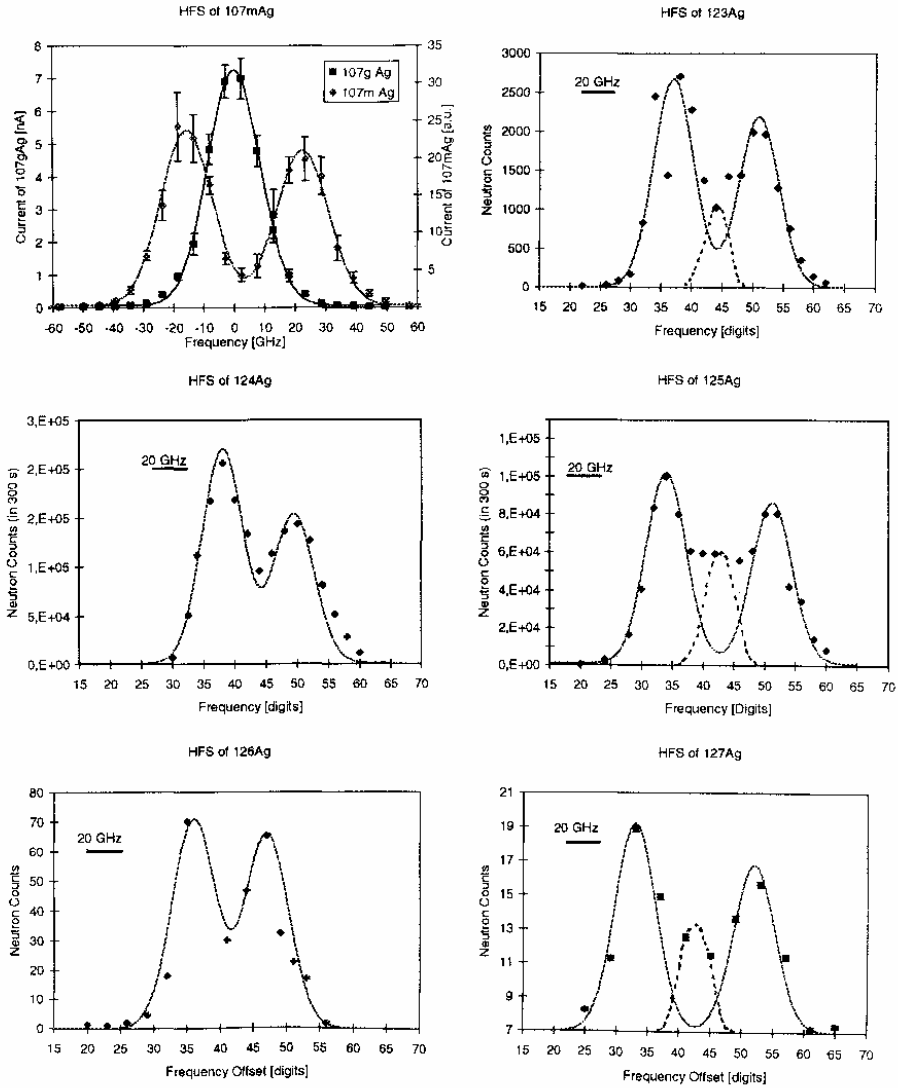,angle=0,width=12.9cm}}
\caption{Laser frequency scans for $^{107}$Ag and $^{123-127}$Ag 
using beam current or $\beta$dn-counting, respectively 
{\protect{\cite{ref:Kratz98,KRAENAM98,ref:Seb98}}}. The global 
distributions of the short-lived radioactive isotopes are consistent with 
the typical HF-splitting of a $\pi$g$_{9/2}$ configuration. 
The odd-mass nuclei show in addition -- as in the case of stable $^{107}$Ag --
a weak central peak indicating the $\pi$p$_{1/2}$ component. } 
\end{figure}
    
The laser response to HF-splitting was calibrated on the stable 
J$^{\pi}$=1/2$^-$ $^{107}$Ag whose magnetic moment is small ($\mu$=-0.11~n.m.) 
and on the radioactive J$^{\pi}$=7/2$^+$ isomer that has a large moment of 
$\mu$=4.4~n.m. (see upper left part of Fig.~8, and Sebastian et al. 
\cite{ref:Seb98}). Subsequently, the $\beta$dn-activities were measured as a 
function of laser frequency for $^{122}$Ag up to $^{127}$Ag (see Fig. 8). 
All Ag isotopes show -- apart from the expected small isotope shift -- the 
same global pattern, where the moment of 
the odd g$_{9/2}$-proton is the dominant source of HF-splitting. 

For $^{122}$Ag, previous decay-studies had already indicated 
the presence of both a low- and a high-spin isomer \cite{ref:Zam95}. And 
indeed, apart from the highly split peak that can be associated with a 
($\pi$g$_{9/2}$--$\nu$h$_{11/2}$)$_{{9}^{-}}$ configuration, another narrow 
peak shows up in the center that could arise from the 
($\pi$p$_{1/2}$--$\nu$d$_{3/2}$)$_{{1}^{-}}$ configuration (see Fig.~9). 
Subsequently, both 
$\beta$dn- and $\gamma$-decay from $^{122}$Ag was studied   
as a function of laser frequency. As is evident from the partial 
$\gamma$-spectra shown in the lower part of Fig.~9, when the laser is set 
at the center of the frequency scale, $\gamma$-rays from both isomers can 
be observed. However, when the laser frequency is shifted away from the 
center (to 35 units), ionization of the low-spin isomer is suppressed, and 
the $\gamma$-spectrum indicates that only the decay of the high-spin isomer 
is oberved. Similarly, when following the $\beta$dn-decay curves as 
a function of laser frequency, different half-lives are obtained which 
represent varying admixtures of the two isomers. From a preliminary analysis,
we get a half-life of the low-spin isomer of 
T$_{1/2}$$\simeq$550(50)~ms and a value of 
T$_{1/2}$$\simeq$200(50)~ms for the high-spin isomer of $^{122}$Ag 
\cite{KRAENAM98,ref:Kaut98}. This experiment clearly demonstrates 
the sensitivity of ionization to laser frequency and marks the first 
spectroscopic application of this technique to short-lived isomers. 
In the meantime, a gross analysis 
of $\gamma$-spectra taken at different laser-frequency settings confirms
isomerism also for the heavier Ag isotopes up to (at least) $^{126}$Ag.

\begin{figure}
\centerline{\epsfig{file=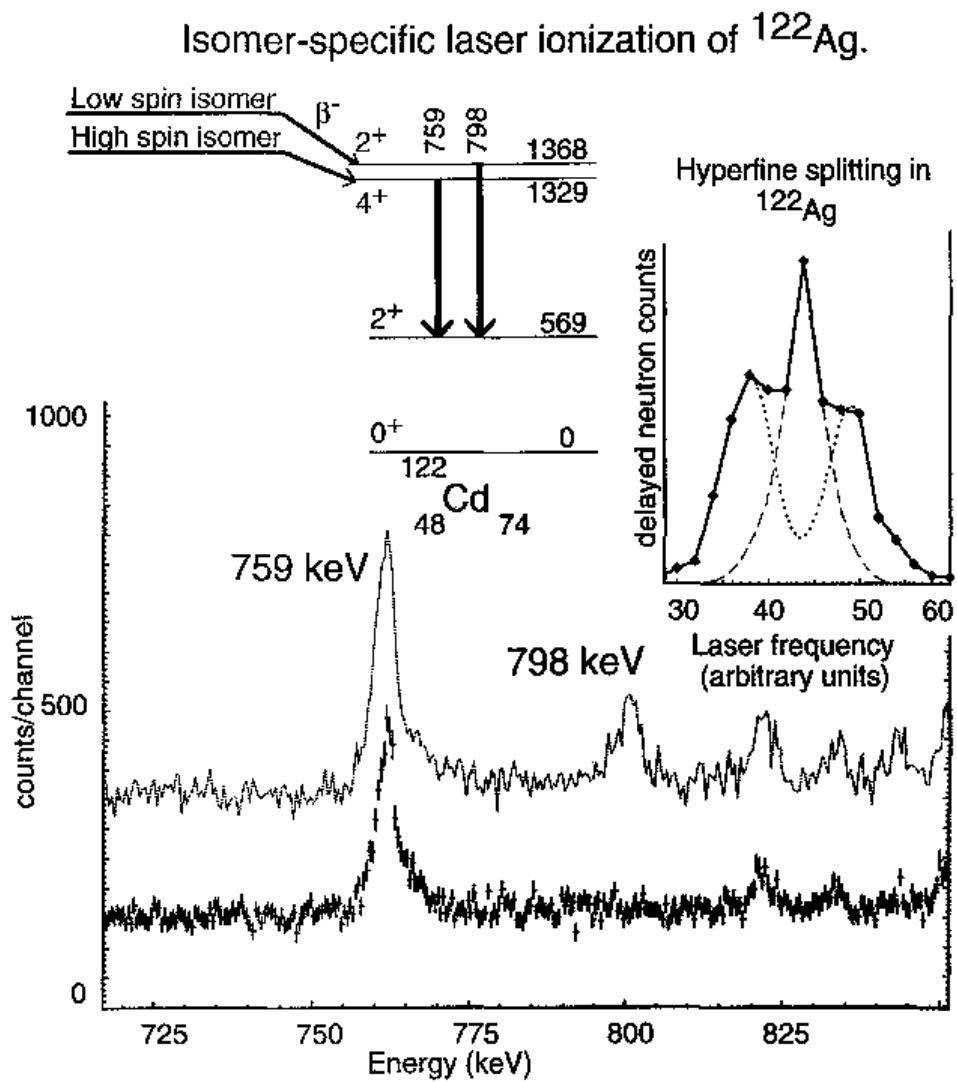,angle=0,width=12.9cm}}
\caption{
Portions of $\gamma$-ray spectra of the decay of 
$^{122}$Ag isomers. 
The upper spectrum was taken with the laser frequency set at the center peak 
(44 units; see insert upper right part), and can be compared with the lower 
spectrum obtained at an off-center frequency (35 units). 
The $\gamma$-ray at 798~keV has been assigned 
to the 2$_{2}^{+}$$\rightarrow$2$_{1}^{+}$ transition (see insert upper left 
part), whereas the line at 759~keV corresponds to the decay of the 
4$_{1}^{+}$ level to the 2$_{1}^{+}$state \protect\cite{ref:Zam95}.
}
\end{figure}

An important nuclear-physics quantity for r-process nucleosynthesis 
calculations (in 
particular for the r-matter flow through the 
A$\simeq$130 region and the N$_{r,\odot}$ peak shape) is the 
$\beta$-decay half-life of the 
N=82 waiting-point isotope $^{129}$Ag, situated just below $^{130}$Cd. 
QRPA calculations using different Q$_{\beta}$-values and 
SP wave functions resulted in decay rates between 15~ms 
and 170~ms \cite{kratz88,kratzea93,ref:MNKz97}. This had left 
large uncertainties in the reproduction of the A$\simeq$130 peak  
(see Fig. 1 in \cite{ref:Kratz98}, and also Fig. 4 in \cite{ref:krat95}).   

Initial attempts \cite{ref:Fedo95} to observe the $\beta$dn-decay 
of $^{129}$Ag, performed with the (broad-band) lasers centered with respect 
to the mean frequency for the stable $\pi$p$_{1/2}$ $^{107}$Ag, had failed. 
After the observation of enhanced ionization 
of the $\pi$g$_{9/2}$ level at an off-center frequency, a new experiment 
was performed with the laser setting at the left $\pi$g$_{9/2}$ peak value 
observed for $^{127}$Ag (see lowest right part of Fig.~8). 
This approach together with the 
microgating procedure \cite{ref:Jad97} mentioned above, finally permitted 
the unambiguous identification of the $\beta$dn-decay from 
$\pi$g$_{9/2}$ $^{129g}$Ag (see Fig.~10).  The half-life of 
46$^{+5}_{-9}$~ms is in very good agreement with the recent QRPA prediction 
of 47~ms \cite{ref:MNKz97}, but is lower than our old 'waiting-point 
requirement' of about 130~ms \cite{kratzea93,ref:krat95}. However, that 
estimate was based on the earlier half-life measurement of 
T$_{1/2}$$\simeq$195 ms for $^{130}$Cd, and 
on the older r-process residuals N$_{r,\odot}$ of \cite{kaepp88}. However, for 
a non-equilibrium r-process, or after the breakdown of the $\beta$-flow 
equilibrium, also the T$_{1/2}$ contribution from the $\pi$p$_{1/2}$ 
$^{129}$Ag isomer would be of importance for the `stellar' half-life.
With the QRPA model of \cite{moran90},
it is possible to calculate the GT decay for such an isomer 
to T$_{1/2}$$\simeq$320~ms. When including an estimate for the expected 
first-forbidden strength (extrapolated 
from the decay of the J$^{\pi}$=1/2$^-$ isomer in isotonic $^{131}$In), a 
minimum value of T$_{1/2}$$\simeq$125~ms is suggested. And, indeed, a careful 
re-examination and comparison of the A=129 $\beta$dn-decay curve taken with the 
laser at central frequency with the pure $^{129}$In curve from a 
laser-off run, gave a first indication of a weak, 'longer-lived' 
$^{129}$Ag $\beta$dn-component with a half-life of roughly 160~ms. 

\begin{figure}
\centerline{\epsfig{file=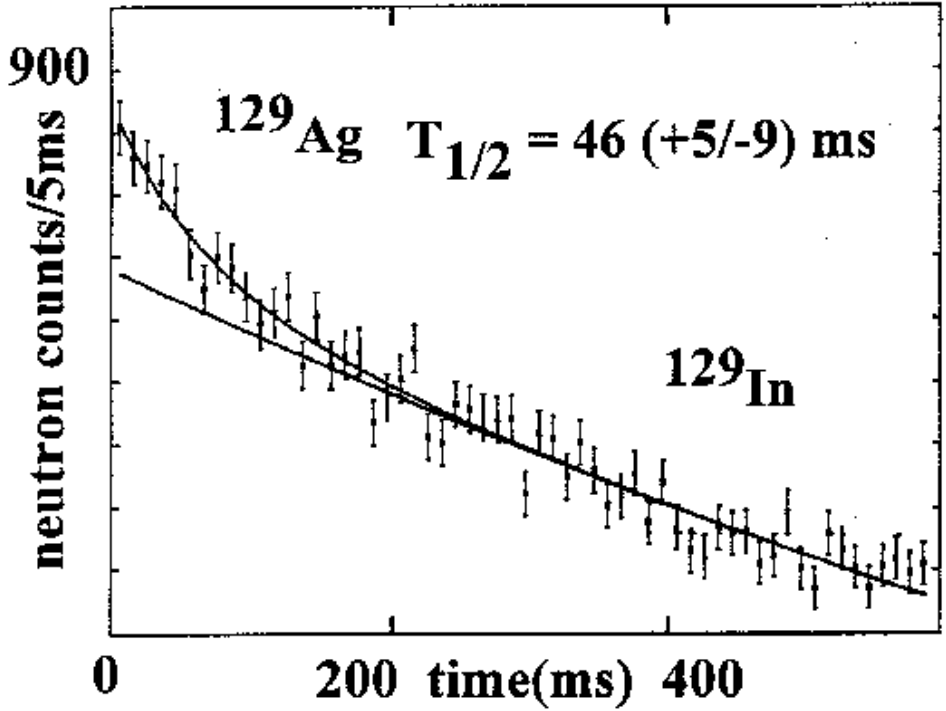,angle=0,width=12.9cm}}
\caption{
$\beta$dn-multiscaling curve for A=129. On top of the isobar activities from 
the two known $^{129}$In isomers, which are due to surface ionization, the
laser-ionized $^{129}$Ag $\pi$g$_{9/2}$ g.s.-decay is shown.
}
\end{figure}

Now, it will be important to use isomer-specific laser ionization in 
combination with isobar separation at ISOLDE-HRS to ascertain the existence of 
this $\pi$p$_{1/2}$ isomer. The astrophysical half-life of $^{129}$Ag for 
non-equilibrium conditions, expected to be a mixture of the g.s and isomer 
values, might then lead to a better understanding of the 'bottle-neck' 
behaviour and  shape of the A$\simeq$130 N$_{r,\odot}$ peak. 
Based on our present knowledge on level systematics in the $^{132}$Sn 
region, neutron-capture $\gamma$-decay from S$_n$$\simeq$5.6 MeV in 
$^{129}$Ag would populate the $\pi$p$_{1/2}$ isomer to roughly 35$\%$ and 
the $\pi$g$_{9/2}$ ground state to about 65 $\%$ \cite{rausch98}. 
This would result in an average stellar T$_{1/2}$$\simeq$80 ms.
However, already within the waiting-point approximation requesting only the 
T$_{1/2}$ of $^{129g}$Ag,
an excellent reproduction of the shape of the N$_{r,\odot}$ peak from A=126 
to A=133 is obtained. From this result, we 
conclude that the effect of neutrino-processing of the N$_{r,prog}$ during
freeze-out is considerably smaller for the A$\simeq$130 peak region than 
recently postulated by Qian et al. \cite{qian98},
and that the main effect is rather due to $\beta$dn-branching during the first 
100 ms of the freeze-out. 

In addition to the study of gross $\beta$-decay properties of neutron-rich 
Ag isotopes, also the level systematics of the Cd-daughters has been 
extended up to the r-process path \cite{ref:Kaut96,ref:Kaut98,KRAENAM98}.
First, an alternate source of nuclear structure for neutron-rich isotopes 
is the investigation of correlated $\gamma$-emission of 
complementary fragments of spontaneous-fission systems. Recent data 
from such studies have revealed surprisingly low 2$^+$ to 6$^+$ energies 
in $^{134}$Sn (only two neutrons beyond double-magic $^{132}$Sn), that 
suggest a reduction of the effects of pairing in very neutron-rich Sn 
nuclides \cite{Sn-134}. For the Cd isotopes, however, such fission data 
extend only up to $^{122}$Cd. Our studies at ISOLDE have now permitted 
even-even Cd structures to be extended six neutrons further out, up to 
N=80 $^{128}$Cd.

The new results are shown in Fig.~11, together with known E(2$^+$) 
and E(4$^+$)/E(2$^+$) level systematics of neighbouring even-Z elements. The 
E(4$^+$)/E(2$^+$) ratio for the 4n-hole nuclide $^{126}$Cd is 2.25, 
a value almost unchanged relative to that for the 6n-hole isotope 
$^{124}$Cd. Already this static ratio is in contrast to the larger reduction 
observed for the Z=52 isotones $^{128}$Te and $^{130}$Te. For the new 2n-hole 
nuclide $^{128}$Cd, the energy of the first 2$^+$ state at 645~keV is even 
{\bf lower} than the E(2$^+$) of 657~keV in $^{126}$Cd, and also the 
E(4$^+$)/E(2$^+$) ratio of 2.22 is only insignificantly smaller than 
those in $^{124}$Cd and $^{126}$Cd. As can be seen from the right side of 
Fig.~11, this trend for the Z=48 isotopes clearly deviates from that 
of the Z=50 (Sn), 52 (Te) and 54 (Xe) isotones. This situation is similar to 
the Hg isotopes below N=126, where the E(2$^+$) in the 2n-hole nuclide 
$^{204}_{\phantom{2}80}$Hg$_{124}$ is with 436.6 keV also slightly lower than 
the E(2$^+$)=439.6 keV in the 4n-hole isotope 
$^{202}_{\phantom{2}80}$Hg$_{122}$.
For the ee-Cd nuclides this indicates an extension of the well-known vibrational
character of the lighter isotopes up to N=82, indicating a weakening of the 
spherical shell strength below $^{132}$Sn. For A=130, we do have an 
indication for two weak $\gamma$-lines at 957 keV and 1395 keV with a tentative
assignment to the 4$^+$ and 2$^+$ levels in $^{130}$Cd. If confirmed,  this
would result in an E(4$^+$)/E(2$^+$) ratio of 1.69.

\begin{figure}
\centerline{\epsfig{file=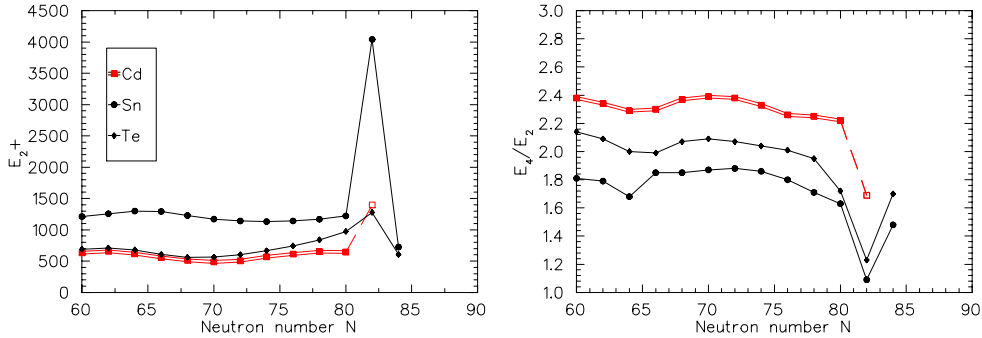,angle=90,width=12.9cm}}
\caption{Systematics of the first 2$^+$ levels in neutron-rich $_{48}$Cd 
to $_{52}$Te isotopes (left part); and E(4$^+$)/E(2$^+$) ratios (right part).}
\end{figure}

As the above results were accomplished with LIS and GPS-ISOLDE, again 
there is promise of additional data with HRS. In particular, a study of 
the decay of isobaric `clean' $^{130}$Ag should permit the identification 
of excited states of N=82 $^{130}$Cd up to the 8$^+$ level in which two aligned 
g$_{9/2}$ proton holes are coupled to double-magic $^{132}$Sn. As the 
structure of $^{98}$Cd, which has the same two g$_{9/2}$ aligned protons 
coupled to the double-magic $^{100}$Sn core, is known, a direct comparison 
is already available for such a measurment at N=82. There also 
exists the possibilty of the presence of a low-spin $\pi$p$_{1/2}$ isomer 
in $^{131}$Ag that should populate the $\nu$p$_{3/2}$ state in the 
$^{131}$Cd-daughter. As has been mentioned previously, one possible effect 
of large neutron excess in nuclei is the lowering of the low-j orbitals.
Hence, an HRS study of $^{131}$Ag decay could demonstrate if the 854-keV gap 
between  the $\nu$p$_{3/2}$ first excited state and the $\nu$f$_{7/2}$ g.s. 
in $^{133}$Sn$_{83}$ has narrowed with the removal of two protons 
in the N=83 Ag isotone. 

Under development and testing at ISOLDE is a scheme for laser ionization of
Sn nuclides \cite{ref:Mish93}. If past experience is any guide, again, it 
should be possible 
with GPS and HRS to produce and study the decay of at least six Sn isotopes 
beyond $^{134}$Sn, the heaviest Sn nuclide for  which some nuclear properties 
are known. $^{140}$Sn is a nucleus with 50 protons and 90 neutrons,
a neutron/proton ratio of 1.8. It would be the most neutron-rich nuclide 
studied beyond the sd shell; and if there are to be strong divergences from
expected nuclear properties because of the high neutron/proton ratio, the
properties of this nuclide should reveal such signatures. Moreover, these
heavy Sn nuclides lie directly in the path of the r-process, and the degree to
which they capture neutrons prior to undergoing $\beta$-decay is a measure of
how quickly the r-process flow can be established beyond the waiting-point
nuclides at N=82. At this point, all nucleosynthesis calculations use 
predicted T$_{1/2}$ values which are presumably too long.

Given the recent, highly interesting results in the $^{132}$Sn region, one is 
tempted to re-inspect existing data 
for additional 'hidden' or so far unrecognized signatures of shell quenching
at N=82. And indeed, first qualitative evidence for such a 
phenomenon came already from a comparison of the old T$_{1/2}$ 
measurement of $^{130}$Cd$_{82}$ \cite{ref:Kratz86}
with predictions from the original QRPA model of M\"oller 
and Randrup \cite{moran90}. When applying 
Q$_\beta$-values from the 'unquenched' mass models FRDM \cite{ref:MNKz97} and
ETFSI-1 \cite{aboussir95}, theoretical T$_{1/2}$ values for GT-decay 
of 1.12~s and 674~ms
are obtained, respectively. However, with the 'quenched' mass formulae HFB/SkP
\cite{dob96} and ETFSI-Q \cite{pearson96}, shorter values of 246~ms and
364~ms, respectively, are derived which are in better agreement with 
experiment. Another, so far unrecognized indication is given by the 
measured masses
in the $^{132}$Sn region \cite{ref:Audi97,ref:Fog98}. As can be seen from 
Fig.~12, between the sequence of $_{50}$Sn and $_{48}$Cd isotopes
there is a significant change in the trend of the experimental and theoretical
mass differences  (normalized to the FRDM predictions, M$_i$-M$_{FRDM}$). 
Clearly, for the Cd isotopic chain the best agreement with the measured 
masses is obtained with  the 'quenched' ETFSI-Q. 

\begin{figure}
\centerline{\epsfig{file=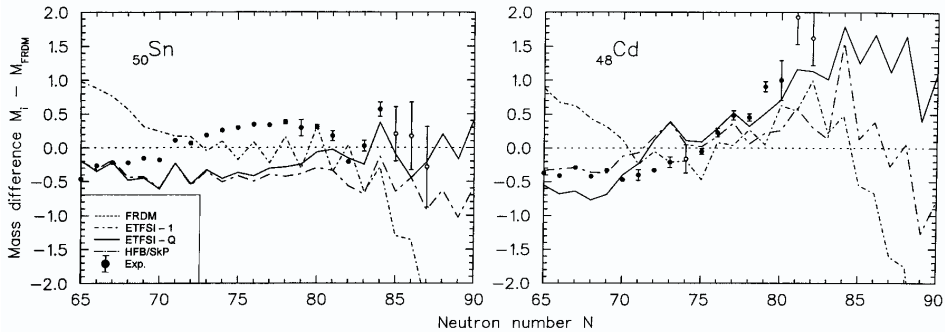,width=12.9cm}}
\caption{
Experimental and theoretical mass differences for $_{50}$Sn and $_{48}$Cd
isotopes normalized to FRDM values (M$_i$--M$_{FRDM}$)
}
\end{figure}

All these signatures can, of course, only be taken as first evidence for a 
'quenching' of the spherical N=82 shell below $^{132}$Sn. Certainly, much more 
experimental data, including nuclear masses (and, as e.g. shown in Fig.~4 
of \cite{chen95}, in particular S$_n$ 
values of N=81 and 83 isotones), single-neutron levels, as well as 
spectroscopic factors of transfer and pickup reactions on radioactive 
isotopes around $^{132}$Sn, are necessary to quantify the effect of shell 
quenching. As will be shown in the next section, a vanishing of the classical, 
spherical N=82 and 126 shell closures would not only be an interesting new 
nuclear-structure phenomenon close to the neutron drip-line, but may also 
have important consequences for r-process nucleosynthesis up to the Th, U, 
Pu cosmochronometers, and maybe even beyond (see, e.g. \cite{cowan99,westphal}).

\section{Solar r-Process Abundances and Astrophysical Sites}

\subsection{Reproduction of the global N$_{r,\odot}$ pattern} 

Mainly as a result of the improved nuclear-physics input, our r-process
parameter studies have revealed that the entire isotopic  N$_{r,\odot}$ 
pattern cannot be reproduced by assuming a global steady flow. Instead, even 
in a single SN II event it requires a
superposition of a multitude of components (minimum three) with different
neutron densities or equivalenty different S$_n$'s, characterizing different
r-process paths and time scales. In our approach, the weighting  of the
individual r-components is naturally given by the $\beta$-decay half-lives of
the three waiting-point nuclei $^{80}$Zn,  $^{130}$Cd and  $^{195}$Tm which
represent the (main) progenitors of the stable isobars  $^{80}$Se,  $^{130}$Te,
 $^{195}$Pt  situated at the top of the respective  N$_{r,\odot}$ peaks.
That this approach  has, indeed,  a nearly one-to-one mapping
to `realistic' astrophysical environments with time variations has been
discussed in some detail in \cite{kratzea93,chen95,freiburghaus99}.
In Fig.~13, we show global  N$_{r}$  distributions from a superpostion of
sixteen n$_n-\tau$ components for two versions of the ETFSI
nuclear mass model (EFTSI-1 \cite{aboussir95} and ETFSI-Q \cite{pearson96}). 
In both  cases, identical conditions for the stellar
parameters were used. Within this framework, the successful reproduction
of the position and relative height of the N$_{r,\odot}$ peaks as well as the
remaining defiencies have been interpreted as signatures of nuclear structure 
near the neutron drip-line \cite{kratzea93,chen95,ref:Kratz98,pfeiffer97}. 
It should be mentioned at this point, that this interpretation is in
contradiction with the conclusions of other authors (see, e.g. 
\cite{bouquelle96}) who prefer a mathematical `Fourier-type' analysis, 
assuming that the N$_{r,\odot}$ pattern represents a `multi-SN-event'
composition.
             
\begin{figure}  
\centerline{\epsfig{file=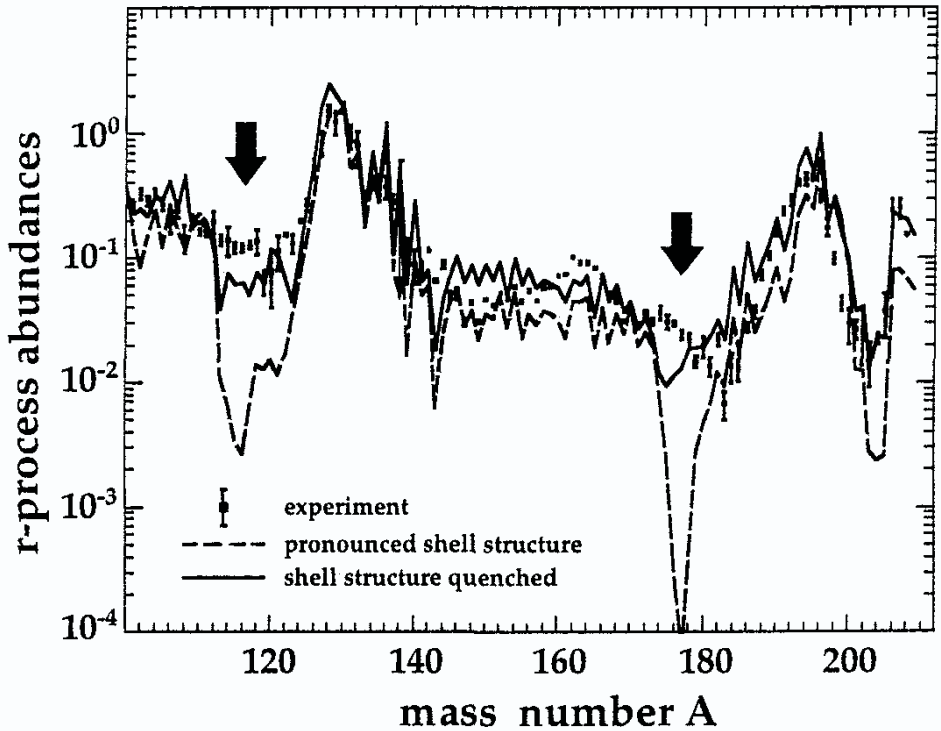,angle=0,width=12.9cm}}
\caption{
Global r-abundance fits for the ETFSI-1 mass model \cite{aboussir95} with
pronounced shell structure and the ETFSI-Q formula \cite{pearson96} with
quenched shell gaps. For discussions, see text and 
\cite{pfeiffer97,ref:Kratz98}.}
\end{figure}

As can be seen from Fig.~13 (dotted fit-curve), with the ETFSI-1 mass model,
apart from pronounced A$\simeq$115 and
A$\simeq$175 abundance troughs, too little r-process material is observed  in
the whole region beyond the A$\simeq$130 peak. The solution to the A$\simeq$115
r-abundance deficiency clearly lies in the prediction of nuclear properties
along the r-process path in this mass range, which have their origin in
an overestimation of the N=82 shell strength below $^{132}$Sn (for a more
detailed discussion see \cite{kratzea93}). Analogous deficiencies in describing
shape-transitions around N$\simeq$104 midshell and the N=126 shell strength seem
to be the origin of the deep A$\simeq$175 r-abundance trough.

It has therefore been suggested   
\cite{kratzea93,thielemann94a,ref:krat95,chen95,pfeiffer97}, 
that the trough problem might be resolved if
for very neutron-rich nuclei the N=82 and 126 shell gaps are less pronounced
than predicted by the global mass models \cite{moeller95,aboussir95} used so
far in r-process calculations. A weakening of sperical shells near stability
with increasing isospin, resulting in a gradual setting in of collectivity, is
well established for the N=20 and 28 shells
(see, e.g. \cite{ref:orr91,ref:sorl93,ref:wern94}),
and is also indicated at N=50 \cite{ref:krat88,kratzea93}.
Quenching of spherical shells at N=82 and 126 has recently been predicted by 
HFB calculations \cite{dobaczewski94,doba96,brown98}, and the infinite nuclear
matter model \cite{satp98}.

In order to check our predictions, in a first step  we have replaced the 
FRDM masses by the  S$_n$'s from the spherical HFB/SkP model \cite{dob96} 
locally in the vicinity of the N=82 shell closure. With this, we could indeed 
observe a considerable improvement of the
calculated r-abundances in the `pathological' A$\simeq$115 region (for a more
detailed discussion, see \cite{chen95} and Figs.~2 and 3 therein).
Motivated by this success, in a next step we have replaced the spherical 
HFB/SkP masses  by the global ETFSI-Q model which takes into account both
deformation and shell quenching \cite{pearson96}.
The full curve in Fig.~13 shows the result of a multicomponent fit with the
latter masses. When compared to the N$_r$ curve obtained with the ETFSI-1
formula, a considerable improvement of the overall fit is observed.
In particular the prominent abundance troughs in the A$\simeq$115 and 175
regions are eliminated to a large extent.  

Hence, a `quenched' mass model for r-abundance calculations beyond
A$\simeq$110 seems to be highly recommended, especially if predictions for the
A$\ge$200 mass region are required. Of special interest in this context is the
good agreement for the region between $^{203}$Tl and $^{209}$Bi with the recent
N$_{r,\odot}$ evaluation of \cite{beer97}. For $^{209}$Bi, for example, 
it indicates a nearly pure r-process origin of this isotope, which confirms 
that (at least in the frame of the double-pulse s-process model) only a minor 
contribution to this isotope is of s-origin. As a consequence, there is no 
further need for the so-called `strong'
s-process component, which had been introduced originally to account for the Pb
and Bi abundances.

This improvement with the ETFSI-Q masses should also make extrapolations up to
the progenitors of the long-lived actinides $^{232}$Th and $^{235,238}$U more
reliable, in particular since they are for the first time based on an internally
consistent  nuclear-physics input. For galactic-age determinations,
commonly production ratios of cosmochronometers, in particular
$^{232}$Th/$^{238}$U, are applied. In \cite{pfeiffer97}, we have compared the
results derived from different mass models to literature values. Only when
using nuclear masses which  correctly describe the entire N$_{r,\odot}$ pattern,
consistent values for a Galactic age of  10-14 Gyrs could be obtained when 
following the procedure described in \cite{cowtt91}.

\subsection{An ultra-metal-poor, neutron-capture-rich halo star}
 
Recently, Sneden et al.~\cite{sneden96} have determined elemental 
abundances of 17 heavy elements between Ba and Th for the 
ultra-metal-poor, neutron-capture-rich halo star CS 22892--052. In 
Fig.~14, these values are compared to the solar r-element 
distribution and to our ETFSI-Q predictions, after adjustment to the 
solar metallicity. The very good agreement clearly indicates, that the 
heavy elements in this star are of pure r-process origin. With the above 
measured Th abundance and our calculated initial abundance, a lower limit 
for the age of this star can be deduced when normalizing to a stable 
r-only element, e.g.~Eu. From the observed ratio Th/Eu=0.219 in this 
halo star, combined with our ETFSI-Q prediction of Th/Eu=0.480 for the 
initial ratio, a lower limit of its `decay age' of 14.5 Gyrs is obtained 
\cite{cowan99}, which overlaps with recent Galactic ages on the young side 
(see, e.g.~\cite{ref:sala97}). 
This implies that the heavy elements in CS 22892--052 must have been 
synthesized already very early in the Galactic evolution, prior to the 
onset of s-process nucleosynthesis \cite{sneden96,cowan99}. Finally, the 
excellent agreement between the r-element abundances of this first-generation 
star and the global N$_{r,\odot}$ pattern, which represents a summation 
over the whole Galactic evolution, indicates that there is probably a 
{\it unique} scenario for the r-process beyond A$\simeq$130. In the
meantime confirmed by observations of other metal-poor stars
\cite{mcwilliam95,cowan99}, 
this result seems to  validate our deductive approach of 
deducing unique astrophysical constraints on r-process conditions from 
fitting the global N$_{r,\odot}$ abundances.  

\begin{figure}
\centerline{\epsfig{file=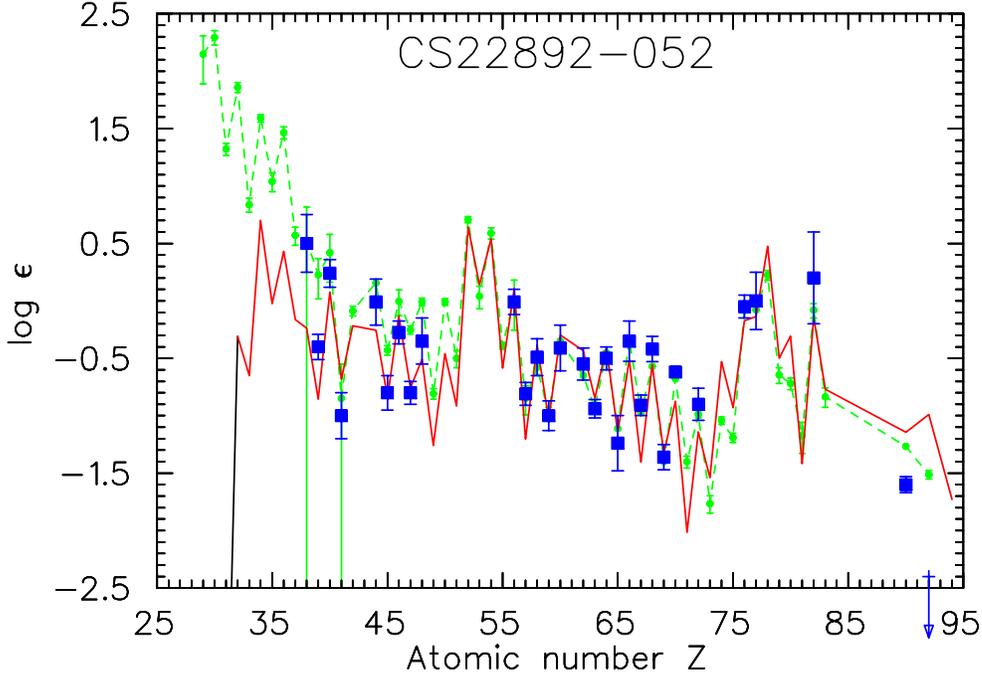,angle=90,width=12.9cm}}
\caption{Elemental r-abundances in units of log $\epsilon$ calculated with 
ETFSI-Q (full line) are compared to the 
respective N$_{r,\odot}$ values (filled circles, connected by dashed line). 
Superimposed are measured abundances (filled squares) from the metal-poor 
halo star CS 22892--052 {\protect \cite{sneden96,cowsne99}}, 
which were normalized to the solar rare-earth values. }
\end{figure}

Only a few months ago, Cowan et al. \cite{cowsne99} have presented new 
element-abundance data for the halo-star CS 22892-052 obtained at the Keck I
high-resolution spectrograph. They partly supercede the earlier analysis of
the same group \cite{sneden96}, and in addition give first results on the 
abundances between $_{41}$Nb and $_{48}$Cd. An interesting feature of the 
latter element pattern is the pronounced odd-even-Z staggering.  All observed
odd-Z elements below Z=50 ($_{39}$Y, $_{41}$Nb, $_{45}$Rh and $_{47}$Ag) are
clearly under-abundant compared to the (metallicity-scaled) solar pattern, 
whereas the even-Z elements ($_{40}$Zr, $_{44}$Ru, $_{46}$Pd and $_{48}$Cd)
are -- within the given errors -- only slightly under-abundant (see Fig.~14). 
These new `low-Z' data seem to support the recent evidence for a 
two-source nature of the solar r-process, originally concluded from extinct 
radioactivities (1.57$\cdot$10$^7$-y $^{129}$I and 9$\cdot$10$^6$-y $^{182}$Hf) 
in meteorites \cite{wass96}.

Taking advantage of our `site-independent', exponential multicomponent approach
to fit the N$_{r,\odot}$ pattern, we now 
can test within the waiting-point assumption under which stellar conditions 
(mainly neutron density and process duration) the possible two r-processes
have to run, in order to reproduce the `low-Z' (below Z$\simeq$50) and `high-Z'
(beyond Z$\simeq$50) element-abundances of CS 22892-052.

When assuming that the abundances of this extremely old metal-poor halo star
are a living record of the first (few) generation(s) of Galactic nucleosynthesis
\cite{cowan99}, the observed pattern beyond Zr up to Th (the trend of the 
lighter elements up to Zr may be explainable in terms of the weak s-process) 
should most likely be produced by only {\bf one} r-process site. This scenario
(the {\it `main'} r-process) then produces the `low-Z' elements under-abundant
compared to solar, and reaches the solar r-pattern presumably beyond $_{48}$Cd.
Fig.~14 shows a fit to the recent CS 22892-052 data, using our standard 
multicomponent approach \cite{kratzea93,pfeiffer97,cowan99} with the ETFSI-Q
nuclear masses. The trend for the `low-Z' elements -- pronounced odd-even 
staggering, and approaching the solar values with increasing Z -- is nicely
reproduced, and the good overall reproduction of the `high-Z' elements (beyond
$_{56}$Ba) is maintained. When starting from an Fe-group seed, this requires
a {\bf minimum} neutron density of n$_n$$\ge$10$^{23}$ cm$^{-3}$ in order to 
simulate the observed partial `drying-out' of the r-matter flow at low Z.
In terms of radiation entropy within the realistic scenario of the high-entropy
wind in SNII (see sect. 5.3), this would translate to S$\ge$200 k$_B$ for 
Y$_e$=0.45 \cite{freiburghaus99}. As a consequence of the present observations 
for CS 22892-052 (but also for other halo stars) and our calculations, the 
`residuals' at low Z relative to the solar distribution, i.e. N$_{r,\odot}$ -
N$_{r,main}$, will require a separate {\it `weak'} r-process component of yet
unknown stellar site. Again, when assuming an Fe-group seed, our calculations 
can reproduce also the resulting N$_{r,weak}$ pattern up to $_{47}$Ag for 
neutron densities of 10$^{20}$$\le$n$_n$$\le$10$^{22}$ at T$_9$=1.35 and a 
process duration of $\tau$$\simeq$1.5 s. A choice of $_{40}$Zr as seed (in 
order to avoid the delay in the A$\simeq$80 bottle-neck region), would speed up
the low-Z production to $\tau$$\simeq$0.5 s. These are conditions that might 
be reached in explosive shell-burning scenarios (see e.g. \cite{ThArHil79}). 
But, before being able to draw more definite conclusions, considerably more 
work has to be done in both astrophysics (site, seed, n$_n$,$\dots$) and
nuclear physics (masses, $\beta$-decay properties, shell-quenching at N=50).

\subsection{Astrophysical Sites}
After having tested nuclear properties with a site-independent approach
in the previous subsection, we want to return briefly to the possible
astrophysical r-process sites introduced already in section 1.2, i.e. SNII
and neutron-star (NS) mergers.
In the first case, a wind of matter from the neutron star surface
(within seconds after a successful supernova explosion) is driven via neutrinos
streaming out from the still hot neutron star 
\cite{woosley94b,takahashi94,hoffman96,hoffman97,qian96b,meyer98}.
This high entropy `neutrino wind' (corresponding to conditions plotted
in the left part of Fig.~1) is leading to a superposition of ejecta with
varying entropies.
An abundance pattern resulting from a superposition of such high-entropy 
environments, representing $\alpha$-rich freeze-outs of various degrees, is 
shown in Fig.~15 \cite{freiburghaus99}. The high entropies (up to
400 $k_b$/nucleon, responsible for the mass region A$>$110) reproduce
nicely the N$_{r,\odot}$ pattern, with the exception of the mass
region 110--120, where the same conclusions can be drawn as in the
site-independent studies (see sect. 5.1). The trough is the result of the 
nuclear-structure deficiencies of the ETFSI-1 mass model below N=82.
The lower entropies are responsible for the abundances in the mass
region 80--110. We see that the predictions for that region do not
reproduce the solar r-abundances. The reason is that essentially no neutrons
are left after an $\alpha$-rich freeze-out with such entropies, and the
abundance pattern is dictated by $\alpha$-separation rather than 
neutron-separation energies. Thus, explaining the r-process by ejecta of SNII
faces two difficulties: (i) whether the required high entropies 
for reproducing heavy r-process nuclei can really be 
attained in supernova explosions has still to be verified, (ii)
the mass region 80--110 cannot be reproduced adequately.
It has to be seen, for example,  whether the inclusion of non-standard 
neutrino properties \cite{mclaughlin99} can cure both difficulties. Given the 
galactic occurence frequency, SNII would need to eject about 10$^{-5}$ 
M$_\odot$ per event.

\begin{figure}
\centerline{\epsfig{file=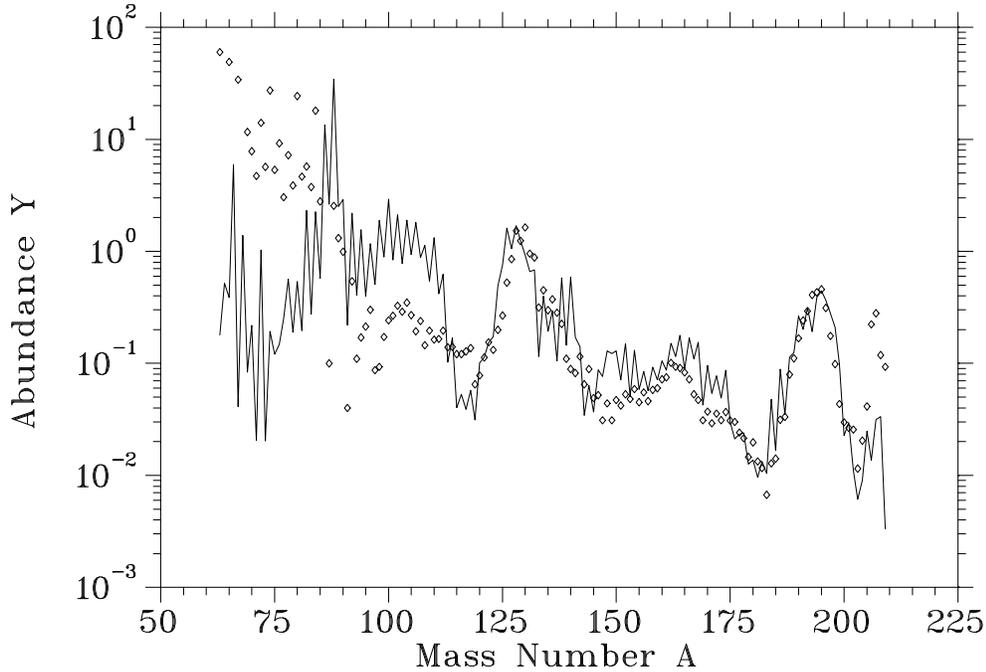,angle=90,width=12.9cm}}
\caption{Fit to solar r-process abundances N$_{r,\odot}$ (open rhombus) with
the ETFSI-1 mass formula {\protect\cite{aboussir95}}, making use of a 
superposition of entropies $g(S)$ from the high-entropy neutrino wind in SNII.
These calculations were performed
with $Y_e=0.45$, but similar results are obtained in the
range $0.30-0.49$, only requiring a scaling of entropy. The trough below
$A\simeq130$ behaves similar to Fig.~13, and can be avoided
by the quenching of shell effects.
The strong deficiencies in the abundance pattern below $A\simeq110$ are due to
the $\alpha$-rich freeze-out and thus related to the astrophysical scenario
rather than to nuclear structure from {\protect\cite{freiburghaus99}}).}
\end{figure}

An alternative site are neutron-star mergers. Interest in a scenario where a 
binary system, consisting of two neutron stars (NS-NS binary), looses energy 
and angular momentum through the emission of gravitational 
waves comes from various sides. Such systems are known to exist; five NS-NS 
binaries have been detected by now \cite{thorsett96}. 
The measured orbital decay gave the first evidence for the existence of 
gravitational radiation \cite{taylor94}. 
Further interest in the inspiral of a NS-NS binary arises from the fact that 
it is the prime candidate  
for a detection by the gravitational wave detector facilities that will come
into operation in the very near future. 
A merger of two NS may also lead to the ejection of neutron-rich 
material, and could be a promising site for the production of r-process 
elements. It is even possible that such mergers account for {\it all} heavy 
r-process matter in the Galaxy \cite{lattimer77,eichler89,rosswog99a}. 
The decompression of cold neutron-star matter has been studied 
\cite{lattimer77,meyer89};
however, a hydrodynamical calculation coupled with 
a complete r-process calculation has not been undertaken, yet. 

Fig.~16 shows the composition of ejecta from a NS merger with an assumed 
superposition of components with $Y_e$=0.08-0.14, as expected when an average
of spin orientation in merger events is taken \cite{rosswog99b}. It is seen 
that the large amount of free neutrons (up to n$_n$$\simeq$10$^{32}$ cm$^{-3}$)
available in such a scenario leads to the build-up of the heaviest elements
and also to fission cycling within very short timescales, while the flow from 
the Fe-group to heavier elements `dries out'. This leads to a complete 
composition void of abundances below the A$\simeq$130 peak. If this tendency is 
confirmed, e.g. by further observations of Z$<$50 elements in very low 
metallicity stars, it would provide strong support for that r-process site, 
but would definitely require an additional astrophysical source which produces 
the bulk of the lighter r-abundances up to A$\simeq$125.

\begin{figure}
\centerline{\epsfig{file=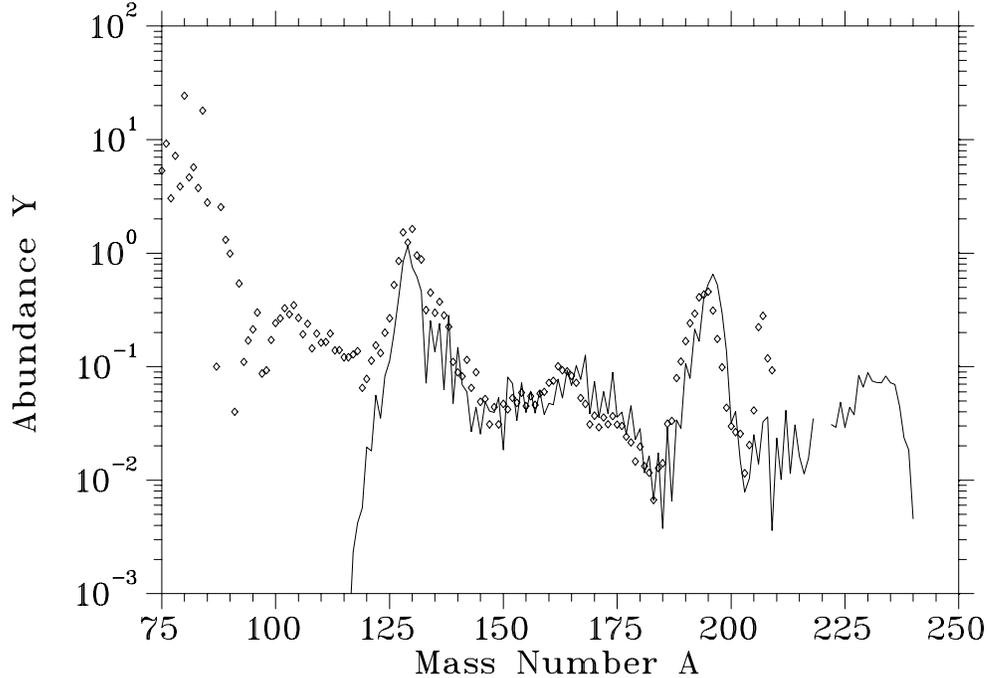,angle=90,width=12.9cm}}
\caption{
Composition of ejecta from a neutron star merger event with an assumed
superposition of components with $Y_e$=0.08-0.14, as expected when an average
of spin orientation in NS merger events is taken. For discussion, see text.
}
\end{figure}

The rate of NS mergers has been estimated to be of the order
$10^{-6}-10^{-5}$y$^{-1}$ per galaxy \cite{eichler89,narayan92};
more recent estimates \cite{vandenheuvel96} 
tend towards the higher end of this range ($8\cdot10^{-6}$y$^{-1}$ per galaxy). 
Hydrodynamic simulations of NS mergers are a formidable
task. Beyond 3D hydrodynamics, it should include
general relativitistic effects, employ a realistic equation of state, contain 
neutrino transport and neutrino cooling, as well as all possible
nuclear reactions \cite{davies94,janka96,ruffert98,baumgarte97}.  
We \cite{rosswog99a,rosswog99b} have performed extensive NS-NS merger 
calculations and find typically ejecta of the order $10^{-2}$M$_\odot$.
This is sufficient to reproduce the solar system r-process abundances,
provided that this matter is converted into
an r-process abundance pattern (which still needs to be shown).

\section{Conclusion}
We have shown in this review which nuclear properties are of prime importance
in the astrophysical r-process. Due to the (partial) equilibrium nature
at high temperatures and neutron densities, the major aspects to explore
are nuclear masses 
and $\beta$-decay properties (half-lives, delayed neutron emission and fission).
The application of recent experimental findings to `site-independent'
r-process calculations has shown that the N$_{r,\odot}$ abundances
actually bear the possibility to test the evolution of nuclear structure
and shell effects as a function of the distance from $\beta$-stability.
In order to put these indications on a more solid footing, the experiments
have to be extended further towards the neutron drip line, which sets
strong requirements for future beam intensities and thus target techniques
as well as experimental methods.
Innovation has played a strong part in establishing a premier role for ISOLDE
in the study of properties of nuclides very far from stability. The development
of the techniques of laser ionization is one example of this kind of 
innovation. A strong lesson from LIS work is that {\it selectivity} is the key
to extending the current results to even more exotic nuclides. Ultimately, the
combination of Z-selective laser ionization, isomer separation via HF-splitting
and the use of the HRS for isobar
separation promises enhanced selectivity.

Astrophysical knowledge is not at all in a more advanced stage. We have
presented two `realistic' sites for the r-process, (i) supernovae of
type II and (ii) neutron star mergers, which represent the two options
to obtain a high neutron to seed-nuclei ratio: high entropies with moderate
$Y_e$ or neutron to proton ratios, or low entropies with small $Y_e$ or
high neutron to proton ratios. 

The first scenario is (as is the SN
explosion mechanism) still riddled with the question whether the required
high entropies of more than 200 $k_B$ per nucleon can be attained at all
and that the lower entropy components give rise to the overproduction of
non-solar r-patterns in the mass range A$\simeq$80--110. Possibly only exoctic
neutrino properties (like sterile neutrinos) might save this environment.

The NS mergers could overcome these problems, if the amount
of ejected matter stays the same in models with general relativity
\cite{rosswog99a}, and a statistical average over
neutron-star spins permits that matter in the range of $Y_e$=0.1
is predominantly ejected \cite{rosswog99b,freiburghaus99b}.
An interesting feature of the second environment is that nuclei with masses 
below the A$\simeq$130 peak are essentially unproduced (see Fig.~16). 

The fact that both scenarios are unable to reproduce the 
N$_{r,\odot}$-`residuals' in the A$\simeq$80--120 region, 
might be related to the increasing evidence of a two-source nature of
the solar r-process, 
which was recently concluded from extinct radioactivities in meteorites 
\cite{wass96} and the low abundances of the odd-Z elements $_{39}$Y to 
$_{47}$Ag measured in CS22892-052 \cite{cowsne99}, and in other
low-metallicity stars (which show for all elements heavier than Ba (Z=56) a 
solar r-process pattern \cite{sneden96,cowan97,cowsne99}). 
Further detailed observations of r-process abundances in low-metallicity stars,
especially for A$<$130 are highly needed for uncovering the nature and stellar 
site of this second ({\it `weak'}) r-process and fixing the time in galactic 
evolution when this nucleosynthesis process sets in.

\subsection*{Acknowledgement}
The authors wish to acknowledge collaboration with many colleagues, in 
particular P. M\"oller,  J. Dobaczewski, J.M. Pearson, B.A. Brown, 
W. Hillebrandt, J.W. Truran, Ch. Freiburghaus, T. Rauscher, S. Rosswog, 
J.J. Cowan, V.N. Fedoseyev, V.I. Mishin, H. Gabelmann, G. Lhersonneau, 
A. W\"ohr, W. B\"ohmer, H.L. Ravn and the ISOLDE Collaboration, and all 
our present and former PhD students and postdocs. Support for this work was 
provided by various grants from the German BMBF and DFG, the Swiss 
Nationalfonds and U.S. DOE.


%
\end{document}